\documentclass[11pt]{spie}  

\usepackage{xcolor}
\usepackage{physics}
\usepackage{amsmath,amsfonts,amssymb}
\usepackage{graphicx} 
\usepackage{caption} 
\usepackage[normalem]{ulem}
\usepackage[numbers]{natbib}
\usepackage[colorlinks=true, allcolors=blue]{hyperref}

\newif\ifshowGraphs
\showGraphstrue  

\let\oldincludegraphics\includegraphics
\renewcommand{\includegraphics}[2][]{%
    \ifshowGraphs
        \oldincludegraphics[#1]{#2}%
    \fi
}

\DeclareMathOperator{\sinc}{sinc}

\newcommand\outerscaleErr{$0.57\pm0.1$}
\newcommand\outerscale{$0.57\pm0.1$}
\newcommand\medianDIQwithDome{$1.12^{\prime\prime}$}
\newcommand\medianDIQwithoutDome{$0.90^{\prime\prime}$}

\definecolor{green2}{rgb}{0.0,0.5,0}

\title{Measurement of dome turbulence intensity from scintillation pattern in the pupil of the telescope}

\author[a]{Victor G. Kornilov}
\author[a]{Matwey V. Kornilov}
\author[a]{Boris S. Safonov}
\author[b]{Anton S. Mironov}
\author[a]{Dmitry V. Cheryasov}
\author[a]{Igor A. Gorbunov}
\author[a]{Ivan A. Strakhov}

\affil[a]{Sternberg Astronomical Institute Lomonosov Moscow State University, Universitetskii prospekt, 13, Moscow, Russia, 119992}
\affil[b]{Faculty of Space Research Lomonosov Moscow State University, Leninskie gory, 1s52, Moscow, Russia, 119234}

\authorinfo{Further author information: (Send correspondence to B.S.)\\E-mail: safonov@sai.msu.ru, Telephone: 7 926 754 3714}
\begin{document}

\maketitle

\begin{abstract}
We present measurements of optical turbulence (OT) power inside the telescope dome using an instrument registering fluctuations of intensity of a bright star at a plane conjugated to $-2$~km below the pupil of the telescope --- Domecam. The measurements were conducted at the 2.5-m telescope of Caucasian Mountain Observatory of Sternberg Astronomical Institute Lomonosov Moscow State University in period 2022--2024 in different ambient conditions. The instrument was validated against Multi Aperture Scintillation Sensor. Using simultaneous observations with a speckle imager mounted at 2.5-m telescope, we demonstrate that dome OT power measured with Domecam can be converted into delivered image quality (DIQ) assuming von K\'{a}rm\'{a}n model with outer scale of \outerscaleErr~m for dome turbulence. Dome turbulence increases median of expected distribution of DIQ for the 2.5-m telescope from \medianDIQwithoutDome to \medianDIQwithDome.
\end{abstract}


\section{Introduction}

Optical turbulence (OT) fundamentally limits the performance of large telescopes operating in seeing-limited mode. It affects two key aspects of astronomical observations: (a) the ability to resolve fine structural details and (b) the detection threshold for faint objects against sky background \citep{Bowen1964}. Beyond image resolution, OT imposes fundamental constraints on photometric precision through scintillation noise \citep{Kornilov2012}. The growing implementation of adaptive optics (AO) systems, which attempt to compensate for OT effects in real-time, further increases the importance of thorough OT characterization, as AO performance is critically dependent on detailed knowledge of turbulence properties.

The optical turbulence (OT) in the free atmosphere is primarily an environmental factor, with limited opportunities for mitigation beyond careful site selection. In contrast, dome OT – originating within the telescope enclosure – can be actively controlled through both design optimization and thermal management. During the design phase, particular attention must be given to minimizing dome OT, while operational measures should focus on maintaining thermal equilibrium between the telescope structure, internal dome air, and external ambient conditions. Crucially, all such mitigation strategies depend fundamentally on accurate quantitative characterization of dome OT intensity, which serves as the essential metric for evaluating both design solutions and thermal regulation performance.

Several methods have been developed to estimate dome OT intensity using both acoustic and optical techniques. Acoustic methods, while practical, require assumptions when converting speed-of-sound variations to refractive index fluctuations \citep{Kornilov1979}, inevitably introducing systematic biases. Optical methods provide more direct and precise measurements for astronomical applications, as their results relate more straightforwardly to optical image formation. However, optical approaches face their own challenge: disentangling the contributions of dome OT from those of free-atmosphere OT. This separation is crucial for accurate characterization of dome-specific effects.

One optical method capable of quantifying dome OT intensity is based on laser beam tip/tilt fluctuations measurements \citep{Bustos2018,Munro2023}. This method employs a 3-4 cm diameter beam propagating between the primary mirror cell and secondary mirror support structure. While the measured tip/tilt variance cannot be directly translated into quantitative dome OT effects on image quality, it can serve as a proxy for dome OT intensity. This design enabled the collection of extensive dome OT statistics across various operational conditions. 

More accurate, but also more expensive in terms of telescope time, approaches involve using the aperture of a telescope and a star as a radiation source. For example, Scintillation Detection and Ranging instrument (SCIDAR) reconstructs the dependence of OT intensity and wind speed on altitude \citep{Vernin1973}. The classical SCIDAR method analyzes spatial cross-correlations of scintillation patterns from binary star components with angular separations of $20-60^{\prime\prime}$. The generalized SCIDAR variant \citep{Avila1997,Klueckers1998} provides sensitivity to near--ground turbulence through optical conjugation of the detection plane several kilometers below the entrance pupil, introducing additional virtual propagation. The dome OT impact on generalized SCIDAR measurements was investigated by \citep{Avila2000}.

Single star SCIDAR measures intensity cross-covariance measured in a plane optically conjugated several kilometers below the entrance pupil using a fast detector with a delay of 10--50~ms. Subsequent analysis of this cross--covariance allows one to resolve turbulent layers by their wind speed and direction \citep{Habib2006,Errazzouki2024}. The authors of \citep{Osborn2023} demonstrated implementation of a similar technique specifically for characterizing dome turbulence using a dedicated Dome Turbulence Monitor (DTM). The DTM utilizes a 20~cm telescope for its feeding optics, mounted adjacent to the main mirror cell of a larger telescope at the European Space Agency's Optical Ground Station in Tenerife, Spain. Atmospheric layers showing zero wind speed were identified as dome (OT) contributions.

The investigation of dome OT influencing factors is particularly relevant for the 2.5-m telescope at the Caucasian Mountain Observatory of Sternberg Astronomical Institute, MSU \citep{Shatsky2020}.  Continuous automatic monitoring with a combined Multi--Aperture Scintillation Sensor and Differential Image Motion Monitor (MASS--DIMM) instrument \citep{Kornilov2007} are conducted at the site of 2.5-m telescope since 2007. The median seeing $\beta$ in period 2007--2009 was measured at $0.93^{\prime\prime}$, with $\beta<0.6^{\prime\prime}$ occurring for 10\% of time. The telescope's optical quality was systematically characterized using a dedicated Shack-Hartmann sensor \citep{Potanin2017}. The analysis demonstrated exceptional optics performance, with 80\% of energy concentrated within a $0.3^{\prime\prime}$ diameter circle --- fully compliant with design specifications and confirming the optical system's excellent quality. The 2.5-m telescope at CMO SAI possesses all the necessary conditions to achieve a delivered image quality (DIQ) $\lesssim0.6^{\prime\prime}$. Such images were indeed obtained, enabling demanding observations \citep{Dodin2019}. 


For implementing the Single Star SCIDAR technique in 2018--2020 we constructed an instrument fed directly by the 2.5-m telescope itself, sampling precisely the same air volume as the telescope's science beam. As this instrument specifically targets dome turbulence measurements, we designated it as a Dome Seeing Camera, or {\it Domecam} for short. In this work we describe the instrument and the method for dome OT intensity estimation. Measurements of dome OT intensity, obtained in 2022--2024 at 2.5-m telescope of SAI are presented and analyzed.



\section{Theory}

\subsection{Auto--covariance of scintillation pattern}
\label{sec:acf}

Consider a two-dimensional uniform grid of identical apertures with spacing $\Delta$. Let 
$W_{nm}({\vb x})$ represent the aperture function for the aperture for the grid node at position $n, m$, where
\begin{equation}
    W_{nm}({\vb x}) = W(x - n \Delta, y - m \Delta),
\end{equation}
with $W({\vb x})$ being the common aperture function. The Fourier transform $\widetilde{W}_{nm}({\vb f})$ of each aperture function satisfies:
\begin{equation}
    \widetilde{W}_{nm}({\vb f}) = \widetilde{W}({\vb f}) \exp\left(-2\pi i \Delta \left(n f_x + m f_y\right)\right),
\end{equation}
where $\widetilde{W}({\vb f})$ is the Fourier transform of $W({\vb x})$. We normalize the aperture functions such that:
\begin{equation}
    \int \vb{d^2 x} W_{nm}({\vb x}) = 1\quad\forall n,m,
\end{equation}
ensuring consistent throughput across all apertures.

Under weak scintillation conditions, the detected flux $I_{nm}$ for the aperture $n,m$ is given by:
\begin{equation}
    I_{nm} = \rho_{nm} \left(1 + 2 \int \vb{d^2 x} W_{nm}({\vb x}) \chi({\vb x})\right),
\end{equation}
where $\chi({\vb x})$ denotes the light-wave amplitude relative fluctuations field. Its Fourier image can be expressed as the following:
\begin{equation}
    \widetilde{\chi}({\vb f}) = \widetilde{\phi}({\vb f}) \sin\left(\pi \lambda z f^2\right),
\end{equation}
where $\widetilde{\chi}({\vb f})$ denotes Fourier image for $\chi({\vb x})$, $\widetilde{\phi}({\vb f})$ is Fourier transform of phase perturbations in a thin layer at the height of $z$, plain font $f$ denotes vector norm $\lVert{\vb f}\rVert$ in the rest of the paper, and $\rho_{nm}$ is an aperture response coefficient for aperture $n,m$. We assume that the aperture grid has finite number of nodes and $\rho_{nm} = 0$ for non-existing (blind) apertures.

Then, the relative flux fluctuation is the following
\begin{equation}
    \label{eq:relative_flux_fluctuation}
    \frac{\Delta I_{nm}}{\left<I_{nm}\right>} \equiv \frac{I_{nm}}{\left<I_{nm}\right>} - 1 = 2 \mathbf{I}_{0^\complement}(\rho_{nm})\int \vb{d^2 f} \widetilde{\phi}({\vb f}) \sin\left(\pi \lambda z f^2 \right) \widetilde{W}({\vb -f}) \exp\left(2\pi i \Delta \left(n f_x + m f_y\right)\right),
\end{equation}
where
\begin{equation}
    \label{eq:pupil_mask}
    \mathbf{I}_{0^\complement}(\rho_{nm}) \equiv
\begin{cases}
0 ~&\text{ if }~ \rho_{nm} = 0, \\
1 ~&\text{ otherwise. }
\end{cases}
\end{equation}

Consider the following quantity:
\begin{equation}
\label{eq:8}
\gamma_{jk} \equiv \sum_{n,m} \left<\frac{\Delta I_{nm}}{\left<I_{nm}\right>} \frac{\Delta I^{*}_{n+j,m+k}}{\left<I^{*}_{n+j,m+k}\right>}\right>.
\end{equation}

Keeping in mind, that
\begin{equation}
\left<\widetilde{\phi}({\vb f}) \widetilde{\phi}^{*}({\vb f'}) \right>= 9.69 \times 10^{-3} \frac{4\pi^2}{\lambda^2} f^{-11/3} C_n^2(z) dz \delta({\vb f} -{\vb f'}),
\end{equation}
we find that
\begin{equation}
\label{eq:10}
    \gamma_{jk} = 9.69 \cdot 10^{-3} \cdot 16 \pi^2 C_n^2(z) dz c_{jk}
    \int \vb{d^2 f}  f^{-\frac{11}{3}}  \frac{\sin^2\left(\pi \lambda z f^2\right)}{\lambda^2} A(-{\vb f}) \exp\left(-2\pi i \Delta \left(j f_x + k f_y\right)\right),
\end{equation}
or the same in the terms of dimensionless variables (see~\citep{Kornilov2021}):
\begin{multline}
     \label{eq:28}
     \gamma_{jk} = 9.69 \cdot 10^{-3} \cdot \frac{16 \pi^2}{\lambda^{\frac{7}{6}}} z^{\frac{5}{6}} C_n^2(z) dz c_{jk} \times \\ \times
     \int \vb{d^2 u} \, u^{-\frac{11}{3}} \sin^2 \left(\pi u^2\right) A\left(\frac{D}{\sqrt{\lambda z}}\vb{u}\right) \exp\left(-2\pi i \frac{\Delta}{\sqrt{\lambda z}} \left(j u_x + k u_y\right)\right).
\end{multline}
We consistently use dimensionless frequencies $\vb u \equiv \sqrt{\lambda z} \, \vb f$ below in the present paper to simplify the results presentation. The proportionality $\gamma_{jk}\sim\lambda^{-7/6}z^{5/6}C_n^2(z)$ was noted before by \citep{Osborn2023}.

In equations~(\ref{eq:10}) and~(\ref{eq:28}), $A(\vb u) \equiv \left|\widetilde{W}(-{\vb u})\right|^2$ is called an aperture filter. The most common aperture filters used in this paper are the filter for an annular aperture with central obscuration $\varepsilon$:
\begin{equation}
    \label{eq:af_annular}
    A({\vb u}) = \left({1-\varepsilon^2}\right)^{-2} \left(\frac{2 J_0\left(\pi u\right)}{\pi u} - \varepsilon^2 \frac{2 J_0\left(\varepsilon \pi u\right)}{\varepsilon \pi u}  \right)^2,
\end{equation}
where $J_0$ denotes the Bessel function, and the unit square aperture filter
\begin{equation}
    \label{eq:af_square}
    A({\vb u}) = \left(\sinc(u_x) \sinc(u_y)\right)^2,
\end{equation}
where $\sinc(x) \equiv \frac{\sin(\pi x)}{\pi x}.$ Note that every aperture filter $A({\vb u})$ obeys the Fourier scaling property, so the aperture size is easily accounted for by substituting ${\vb u}$ with $D\, {\vb u}$ as occurs in equation~(\ref{eq:28}). We refer to $D$ as the aperture scale in the general case. The aperture scale is the diameter for a circular aperture or the size length for a square aperture. When considering a CCD or CMOS detector, it is also reasonable to assume that the aperture scale coincides with the pixel grid step, i.e $D = \Delta$.

We refer to $c_{jk}$ as the pupil transfer coefficient. For example, if $\rho_{nm} = 1$ for $0 \le n < N$, $0 \le m < M$, and $\rho_{nm} = 0$ otherwise; then $c_{jk} = \max \left\{0,  \left(N - j\right) \left(M - k\right) \right\}$.
Thus, $c_{jk}$ represents the total number of pixel pairs separated by $j$ and $k$ steps in each dimension.
The coefficients $\mathbf{I}_{0^\complement}(\rho_{nm})$ can alternatively be represented numerically as a binary matrix. This representation proves particularly convenient when working with the exit pupil images. In such cases, $c_{jk}$ can be efficiently computed numerically using, for instance, the fast Fourier transform algorithm.

\begin{figure}[t!]
\centering
\includegraphics[clip=true, trim={0cm 3cm 0cm 3cm},width=15cm]{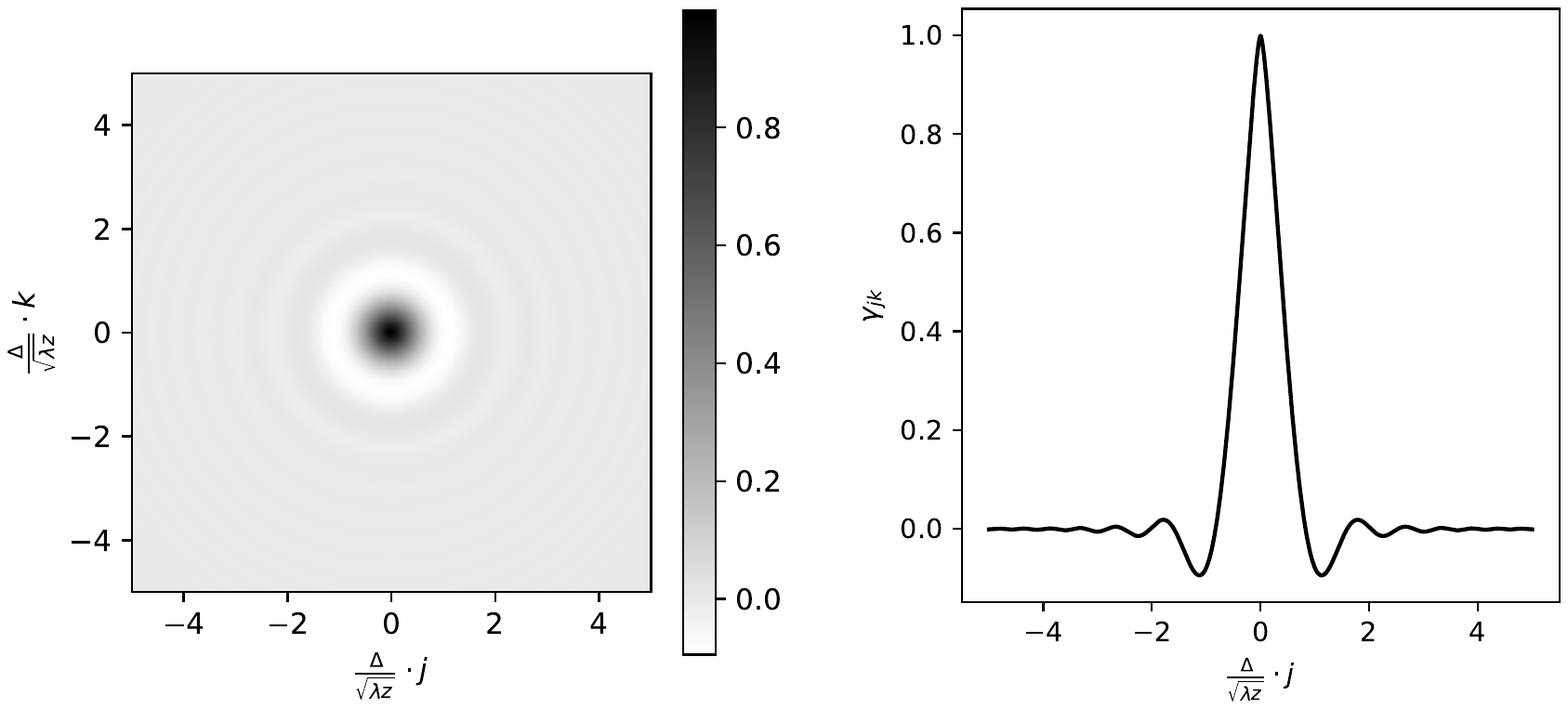}
\caption{
Autocorrelation function for relative flux fluctuation in case of infinitely small aperture.
Left panel: two dimensional density plot.
Right panel: one dimensional cut along symmetry axis.
}
\label{fig:acf_point}
\end{figure}

In case of infinitely small aperture, i.e. $A({\vb u})=1$, equation~(\ref{eq:28}) can be evaluated in the following closed form
\begin{multline}
\label{eq:mono_point_acf}
\gamma_{jk} = 9.69 \cdot 10^{-3} \cdot \frac{32 \pi^3}{\lambda^{\frac{7}{6}}} z^{\frac{5}{6}} C_n^2(z) dz c_{jk} \left(\,_2F_3\left(\left[-\frac{5}{12},\frac{1}{12}\right],\left[\frac{1}{2},\frac{1}{2},1\right],-\zeta^2\right)\right.\\
-\frac{12}{5}\frac{\Gamma\left(\frac{7}{12}\right)}{\sqrt{\pi}\Gamma\left(\frac{11}{12}\right)} \sec\left(\frac{\pi}{12}\right) \csc\left(\frac{\pi}{12}\right) \sin^2\left(\frac{5}{12}\pi\right) \zeta^{5/6}\\
+\left.\frac{5}{3}\csc\left(\frac{\pi}{12}\right)\sin\left(\frac{5}{12}\pi\right)\zeta\,_2F_3\left(\left[\frac{1}{12},\frac{7}{12}\right],\left[1,\frac{3}{2},\frac{3}{2}\right],-\zeta^2\right)\right),
\end{multline}
where $\zeta \equiv \pi {\Delta^2 \left(j^2+k^2\right) } / \left({4 \lambda z}\right)$, and $\,_2F_3$ denotes generalized hypergeometric function.
We have employed the {\tt mpmath}~\citep{mpmath} Python package to accurately evaluate the generalized hypergeometric functions. Fig.~\ref{fig:acf_point} shows a plot of Equation~(\ref{eq:mono_point_acf}), demonstrating the typical behavior of $\gamma_{jk}$.

When the aperture has finite size, i.e. $A({\vb u}) \ne 1$, Equation~(\ref{eq:28}) can be transformed by applying the convolution theorem as follows:
\begin{multline}
    \label{eq:convolution_gamma}
    \gamma_{jk} = 9.69 \cdot 10^{-3} \cdot \frac{16 \pi^2}{\lambda^{\frac{7}{6}}} z^{\frac{5}{6}} C_n^2(z) dz c_{jk} \int \vb{d^2 x'} \int \vb {d^2 v}  A\left(\frac{D}{\Delta} {\vb v}\right) \times \\ \times \exp\left(-2\pi i \left((j - x'_1) v_x + (k - x'_2) v_y\right)\right) \int \vb{d^2 u} \, u^{-\frac{11}{3}} \sin^2 \left(\pi u^2\right) \exp\left(-2\pi i \frac{\Delta}{\sqrt{\lambda z}} \left(x'_1 u_x + x'_2 u_y\right)\right).
\end{multline}
In this case, accounting for the aperture size corresponds to an anisotropic convolution (blurring) of the resulting autocorrelation image. 
The blurring kernel is the inverse Fourier transform of the aperture filter, which equals the autocorrelation of the aperture pupil function. For example, for the aperture filter~(\ref{eq:af_square}), the blurring kernel is simply a square pyramid.
Note that the blurring kernel does not depend on wavelength or altitude, meaning that the scintillation pattern generated by atmospheric optical turbulence located at higher altitudes is less affected by finite aperture size than that from lower altitudes.

Following the same technique as in~\citep{Tokovinin2003} and applying the equivalent wavelength framework from~\citep{Kornilov2021}, we can derive the general equation for the polychromatic case:
\begin{equation}
     \label{eq:12}
     \gamma_{jk} = 9.69 \cdot 10^{-3} \cdot \frac{16 \pi^2}{\lambda_0^{\frac{7}{6}}} z^{\frac{5}{6}} C_n^2(z) dz c_{jk} \int \vb{d^2 u} \, {u}^{-\frac{11}{3}} E(u) A\left(\frac{D}{\sqrt{\lambda_0 z}}\vb{u}\right) \exp\left(-2\pi i \frac{\Delta}{\sqrt{\lambda_0 z}} \left(j u_x + k u_y\right)\right),
\end{equation}
where $E(u)$ is called a dimensionless polychromatic Fresnel filter (or simply spectral filter) and $\lambda_0$ is an equivalent wavelength. Both the spectral filter and equivalent wavelength are typically determined numerically using empirical spectral energy distribution curve. Fig.~\ref{fig:spectral_filter} presents examples of these spectral filters $E(u)$.
Note that $\gamma_{00}$ in equation~(\ref{eq:12}) represents the scintillation index $s^2$, which for symmetric apertures (such as annular aperture~(\ref{eq:af_annular})) takes the following well known form:
\begin{equation}
        \label{eq:scintillation_index}
s^2 = 9.69 \cdot 10^{-3} \cdot \frac{32 \pi^3}{\lambda_0^{\frac{7}{6}}} z^{\frac{5}{6}} C_n^2(z) dz \int d u \, {u}^{-\frac{8}{3}} E(u) A\left(\frac{D}{\sqrt{\lambda_0 z}}u\right).
\end{equation}

In order to avoid solving inverse problems in this work, we follow the approach from~\citep{Kornilov2021} allowing us to obtain estimators for common quantities of interest.
Such quantities could be total OT power on the line of sight,
dome OT power,
or MASS scintillation indices to compare with~(see Appendix~\ref{sec:mass}).

First, let us consider the following linear combination of $\gamma_{jk}$ with arbitrary real coefficients $\omega_{jk}$:
\begin{equation}
    \label{eq:gamma_sum}
    \tilde{s} \equiv \sum_{j,k} \omega_{jk} \gamma_{jk} = 9.69 \cdot 10^{-3} \cdot \frac{16 \pi^2}{\lambda_0^{\frac{7}{6}}} z^{\frac{5}{6}} C_n^2(z) dz \int \vb{d^2 u} \, {u}^{-\frac{11}{3}} E(u) A\left(\frac{D}{\sqrt{\lambda_0 z}}\vb{u}\right) \Omega\left(\frac{\Delta}{\sqrt{\lambda_0 z}} \vb u\right),
\end{equation}
where $\tilde{s}$ is referred to as a synthetic scintillation index and $\Omega({\vb u})$ stands for
\begin{equation}
\label{eq:digital_filter}
\Omega({\vb u}) \equiv \sum_{j,k} \omega_{jk} \exp\left(-2\pi i \left(j u_x + k u_y\right)\right).
\end{equation}
We refer to $\Omega({\vb u})$ as the digital filter and to corresponding $\omega_{jk}$ as the digital filter impulse.
Note that the pupil transfer coefficients $c_{jk}$ disappear from the right-hand side of the equation. This can be formally justified if $\omega_{jk}$ are constrained to be zero when corresponding $c_{jk} = 0$.

Second, for sufficiently large grids, unconstrained $\omega_{jk}$ enables construction of $\Omega({\vb u})$ with useful properties as the following.
Since equation~(\ref{eq:digital_filter}) has the form similar to the function Fourier transform,
we can evaluate the filter impulse $\omega_{jk}$ for digital filter $\Omega({\vb u})$ of our choice using inverse Fourier transform with some limitations in mind of course.
There are two major limitations to consider.
First, filter impulse $\omega_{jk}$ are constrained to be zero for large $j$ and $k$ due to finite grid size which affects the smoothness of function.
Second, $\Omega({\vb u})$ is periodic with a period of $1$ by construction.
The latter implies that the smaller ${u}^{-\frac{11}{3}} E(u)$ for $\left|u_x\right|, \left|u_y\right| \ge \frac{\sqrt{\lambda_0 z}}{2 \Delta}$ the better approximation is. For instance, with $z=1\,\mathrm{km},$ $\lambda_0 = 742\,\mathrm{nm},$ and $\Delta = 11\,\mathrm{mm}$ the right hand-side fraction is something close to $1.2$, which seems to be reasonable limit when looking at Fig.~\ref{fig:spectral_filter}.

Let us consider the following two useful choices of $\Omega({\vb u})$. 
First, we choose $\Omega({\vb u})$ to approximate $K \cdot u^{5/3} A^{-1}(\frac{D}{\Delta}{\vb u})$ as good as possible where $K$ is normalization constant to be defined further.
Note that substituting $\Omega\left(\frac{\Delta}{\sqrt{\lambda_0 z}} \vb u\right) = \left(\frac{\Delta}{\sqrt{\lambda_0 z}}u\right)^{5/3}$ into equation~(\ref{eq:gamma_sum}) eliminates dependence on $z$ which means that final measured value is as simple as
\begin{equation}
\tilde{s} = \int C_n^2(z) dz,
\end{equation}
where $K$ is chosen to compensate the constant before integral and evaluated numerically.
Even though we can get $u^{5/3}$ only for $\left|u_x\right|, \left|u_y\right| < \frac{1}{2}$ due to periodic nature of digital filter $\Omega({\vb u})$, it is sufficient to estimate OT power on the altitudes higher than $z=1\,\mathrm{km}$ with our device configuration.
Sensitivity for OT on the lower altitudes is limited due to periodic nature of $\Omega({\vb u})$ with the full accordance to the physical intuition that scintillation is not sensitive to low altitude OT.
Also note, that in the typical case when $D = \Delta$, division by $A({\bf u})$ is perfectly safe within $\left|u_x\right|, \left|u_y\right| < \frac{1}{2}$ since Equation~(\ref{eq:af_square}) has zeroes only at $u_x, u_y = 1$.

Second, we consider $\Omega({\vb u})$ which is a reasonable approximation for the function
$A'(\frac{D'}{\Delta}{\vb u}) A^{-1}(\frac{D}{\Delta}{\vb u})$,
where $A'$ is an alternative aperture filter and $D'$ is an alternative aperture scale. In this case, equation~(\ref{eq:gamma_sum}) provides a method to estimate the scintillation index for another instrument without requiring profile restoration procedures or solving inverse problems. We employ this property in Appendix~\ref{sec:mass} to compare direct Domecam and MASS measurements between each other.

\subsection{Wind effect}

To develop a comprehensive instrument model, we must account for the wind shear impact on auto-covariance $\gamma^{(t)}_{jk}$ measured with time delay $t$. 

Under famous Taylor's frozen flow hypothesis, wind transport perturbations without distortion in a fixed direction. Consequently,  the Fourier image of light-wave amplitude relative fluctuations field at time $t$ is described by the following equation:
\begin{equation}
    \widetilde{\chi}({\bf f}, t) = \widetilde{\chi}({\bf f}, 0) \exp\left(-2\pi i t \left(w_x(z) f_x + w_y(z) f_y \right)\right),
\end{equation}
where ${\bf w}(z)$ represents the wind shear vector at altitude $z$, and $\widetilde{\chi}({\bf f}, 0)$ denotes the initial field Fourier image. This leads to the generalized form of Equation~(\ref{eq:12}), describing the auto-covariance for a single turbulent layer:
\begin{multline}
     \label{eq:acf_time_delay}
     \gamma^{(t)}_{jk} = 9.69 \cdot 10^{-3} \cdot \frac{16 \pi^2}{\lambda_0^{\frac{7}{6}}} z^{\frac{5}{6}} C_n^2(z) dz c_{jk} \times \\ \times
     \int \vb{d^2 u} \, {u}^{-\frac{11}{3}} E(u) A\left(\frac{D}{\sqrt{\lambda_0 z}}\vb{u}\right) \times \\ \times \exp\left(-2\pi i \frac{\Delta}{\sqrt{\lambda_0 z}} \left(\left(j-\frac{w_x(z) t}{\Delta}\right) u_x + \left(k-\frac{w_y(z) t}{\Delta}\right) u_y\right)\right).
\end{multline}
Please note the structure of Equation~(\ref{eq:acf_time_delay}), it reveals that the auto-covariance pattern is simply translated proportionally to the wind speed in each turbulent layer. 

Due to independence of fluctuation of refractive index in different turbulent layers one can obtain $\gamma^{(t)}_{jk}$ for the whole atmosphere integrating over altitude $z$:
\begin{multline}
     \label{eq:acf_time_delay_z}
     \gamma^{(t)}_{jk} = 9.69 \cdot 10^{-3} \cdot \frac{16 \pi^2}{\lambda_0^{\frac{7}{6}}} c_{jk} \int dz C_n^2(z) z^{\frac{5}{6}} \int \vb{d^2 u} \, {u}^{-\frac{11}{3}} E(u) A\left(\frac{D}{\sqrt{\lambda_0 z}}\vb{u}\right) \times \\ \times \exp\left(-2\pi i \frac{\Delta}{\sqrt{\lambda_0 z}} \left(\left(j-\frac{w_x(z) t}{\Delta}\right) u_x + \left(k-\frac{w_y(z) t}{\Delta}\right) u_y\right)\right).
\end{multline}

Since dome wind speeds are typically much lower than free-atmosphere wind speeds, Equation~(\ref{eq:acf_time_delay_z}) predicts that dome turbulence contributions will dominate the central region of the auto-covariance image $\gamma^{(t)}_{ij}$ for $t>0$. This spatial separation enables discrimination between dome and atmospheric turbulence.

\begin{figure*}[t!]
\centering
\includegraphics[width=15cm]{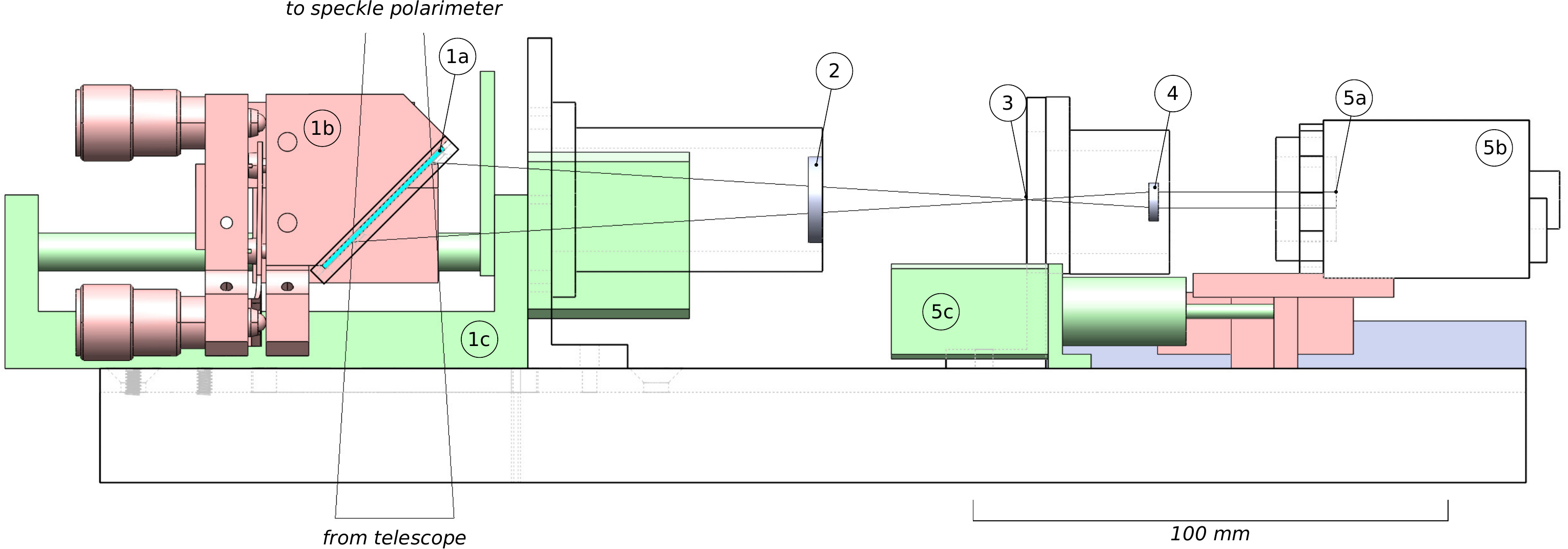}
\caption{Schematic of Domecam (to scale). Translation drives are shown in green, moving parts are highlighted in pink. Key components: (1a) dichroic mirror, (1b) two coordinate kinematic tilt mount of dichroic mirror, (1c) translation drive for the dichroic mirror; (2) filter; (3) field diaphragm; (4) Fabry lens; (5a) focal plane, (5b) detector, (5c) detector translation drive.
\label{fig:scheme}}
\end{figure*}


\begin{figure*}[t!]
\centering
\includegraphics[width=17cm]{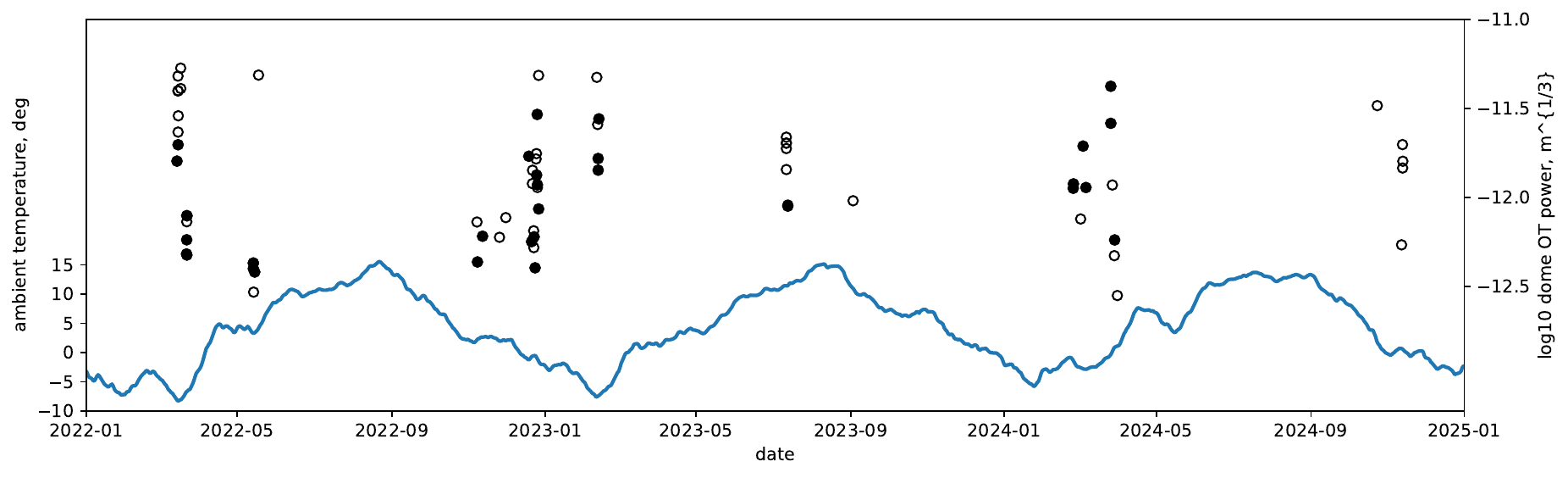}
\caption{Decimal logarithm of dome OT power measured with Domecam is indicated by circles. Filled circles show the cases when simultaneous measurements of DIQ with SPP are available. Daily averaged ambient temperature is shown in below. 
\label{fig:timeline}}
\end{figure*}

\section{Instrument}

The outline of Domecam is displayed in Fig.~\ref{fig:scheme}. The converging ($F/8$) beam from the telescope falls onto a dichroic mirror, which reflects radiation with $\lambda>600$~nm. The reflected beam passes through the KS-19 filter and forms an image of the star at the field diaphragm. The field diaphragm has a diameter of 4~mm, which corresponds to $40^{\prime\prime}$ in the sky. The Fabry lens, installed downstream the focal plane, generates a collimated beam with a diameter of 3.38~mm. The collimated beam forms the exit pupil in the detector.

The detector is a CCD Prosilica GC650. The size of the light--sensitive area is $659\times493$~pixels, with a pixel size of 7.4~$\mu$m. The detector can be moved along the optical axis using a remotely controlled translation drive having effective resolution of $1600$ steps per millimeter. Precise focusing is done by maximizing the sharpness of the outer edge of the pupil using dedicated algorithm \citep{Kornilov2020}. Uniform illumination of the pupil by the twilight sky or the Moon is used for this purpose.

\begin{figure}[t!]
\centering
\includegraphics[width=9cm]{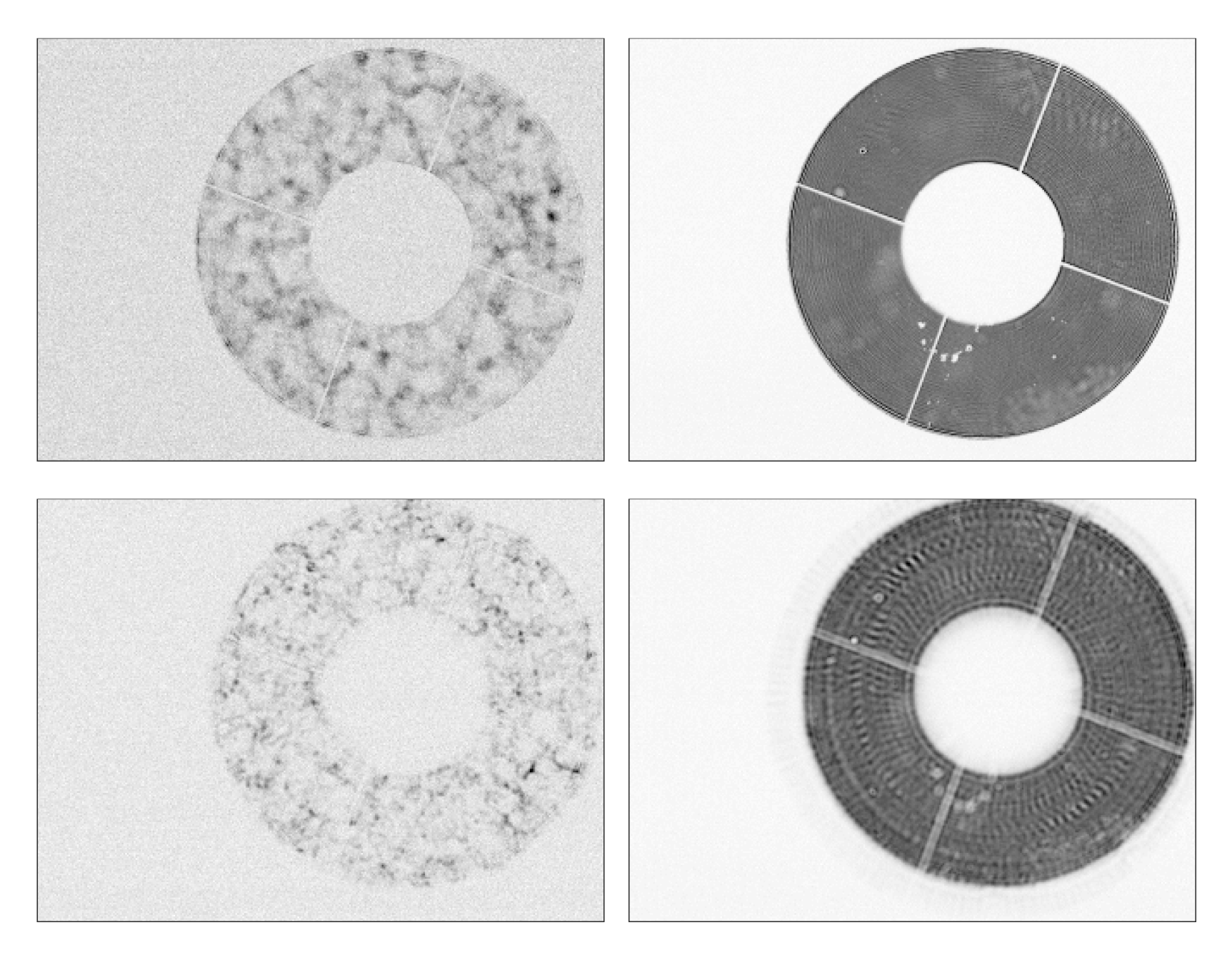}
\caption{Example raw detector frames from Domecam. Left column: single 4 ms exposure frame. Right column: 60 s time-averaged image (6000 frames). Top row: exit pupil plane (conjugation altitude 0~km). Bottom row: $-2$~km conjugation altitude. The radial ripples in lower right image are likely due to the polishing procedure of the main mirror. 
\label{fig:pupils}}
\end{figure}

The plane located 3.65~mm behind the exit pupil is optically conjugated with a plane 2~km below the entrance pupil. Thus, the scheme provides virtual propagation, which is necessary for phase fluctuations generated by dome OT, to develop into amplitude fluctuations, i.e. scintillation, and become detectable. We employ $2\times2$ binning, so the projection of the binned pixel onto the entrance pupil is 1.1~cm. Propagation for 2~km at a characteristic wavelength of radiation $\lambda=650$~nm yields a first Fresnel zone radius of $\sqrt{\lambda z}=3.6$~cm. Hence, the diffraction pattern generated by dome OT is well--sampled. Typical exposures are 2--4~ms, with a frame rate of 100 s$^{-1}$. The series duration is 60~s. Example frames from the detector are shown in Fig.~\ref{fig:pupils}.

Domecam is installed in the Nasmyth focal station N2 (Eastern if the telescope is pointed to the South) of the 2.5-m telescope, between the telescope's mechanical interface and the facility instrument SPeckle Polarimeter (SPP) \citep{Strakhov2023}. During regular astronomical observations, the dichroic mirror (which relays light into Domecam) is moved out of the beam by a remotely controlled translation drive. When the dichroic mirror is placed in the beam, the short--wavelength ($\lambda<600$~nm) radiation is transmitted into the SPP.  This is necessary because the Domecam itself does not have an imaging channel, and the scintillation pattern by itself does not allow monitoring the star's position in the field diaphragm. The SPP provides this imaging channel, enabling the star to be centered in the field diaphragm and tracked during the series acquisition. Additionally, the images recorded by the SPP during Domecam measurements, can be used for independent control of DIQ, see discussion in subsection~\ref{subs:outerscale} of section~\ref{sec:estimation}.  


The instrument is controlled by a dedicated program running in the background on an industrial PC. Domecam is installed at the telescope together with SPP as a single integrated unit. The instrument is available for measurements approximately $\approx80\%$  of the time (for the remaining time the N2 focal station is occupied by another instrument). 

A duty observer performs the observations from the SPP's graphical user interface. To perform measurements, the observer simply needs to point the telescope at the target star and press a single button. Measurements are typically taken after focusing the SPP. Standard procedure of measurement involves obtaining two 1~minute series for 0~km and $-2$~km conjugation altitudes with Domecam and simultaneous data from SPP. 

Measurements with Domecam require pointing of the telescope to a bright star, for this paper we observed stars brighter than $R\approx2.8^m$. Usually the star currently being measured by Multiaperture Scintillation Sensor--Differential Image Motion Monitor (MASS--DIMM, \citep{Kornilov2010}) is selected.  MASS-DIMM is located at the distance of 20~m from the telescope's tower. A single measurement takes 5 minutes, including overheads for pointing and focusing the telescope. In total, 88 measurements were accumulated in 2022--2025, covering different seasons, see Fig.~\ref{fig:timeline}.

\begin{figure*}[t!]
\centering
\includegraphics[width=16cm]{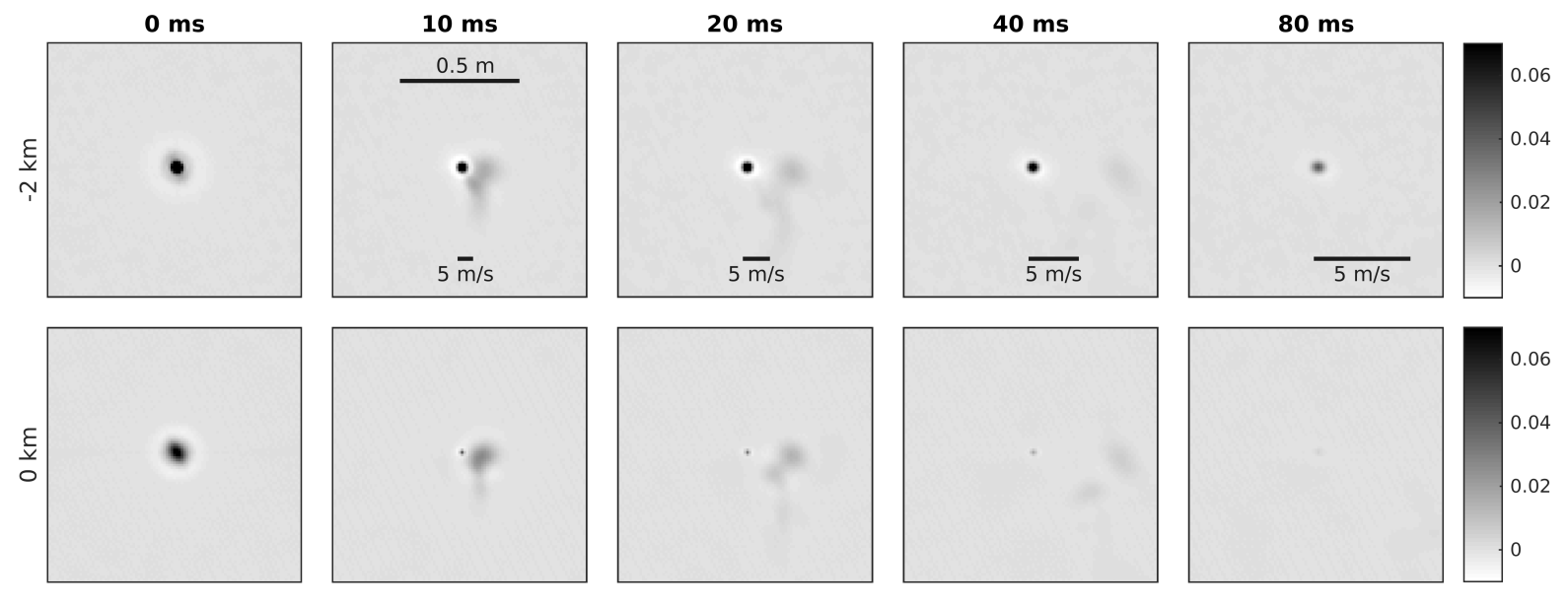}
\caption{Example of auto--covariance measurement, data taken at 25th March 2024, 19:25:59 UT. Columns correspond to different time delays, rows --- to different conjugation altitudes. Field of view is the same for all panels: 1.1~m. Short line in the bottom of top--row panels shows wind speed scale.
}
\label{fig:ACF_example}
\end{figure*}

\section{Estimation of auto-covariance}
The CCD detector provides two fundamental data products for both real and virtual pupils: a scintillation data cube and a mean camera bias frame. The mean bias frame serves to compensate for the non-uniform bias inherent to all CCD detectors. By subtracting this bias frame from each layer of the scintillation cube, we obtain the debiased data cube. This debiased cube forms the foundation for all subsequent processing steps, which are applied consistently to both real and virtual pupil data.

A crucial preliminary step involves estimating the pupil mask $\mathbf{I}_{0^\complement}(\rho_{nm})$ defined in Equation (\ref{eq:pupil_mask}), requiring identification of all blind pixels. This estimation process begins with accumulation of the data cube along the time, followed by application of the threshold detection algorithm from \citep{Otsu1979}. The result is a binary pupil mask (a frame of zeros and ones) that fulfills two essential functions in our analysis. First, the pupil mask enables proper calculation of relative flux fluctuations according to Equation (\ref{eq:relative_flux_fluctuation}) by effectively masking blind pixels. When multiplied with each frame of the debiased cube (element-wise multiplication), it produces what we term the {\it clean} data cube, where all blind pixels are exactly zero and contribute no noise. Second, the pupil mask provides the pupil transfer coefficients $c_{jk}$ through its auto-covariance, as originally introduced in Equation (\ref{eq:10}).

Following the initial cleaning steps, we perform per-frame flux normalization on the clean data to account for temporal variations in atmospheric extinction. This normalization involves calculating the total flux across all pixels for each frame, then dividing each pixel value by integrated flux. While this procedure effectively removes extinction effects, it introduces spatially-correlated fluctuations due to scintillation noise in the 2.5~m telescope aperture, as discussed in \citep{Kornilov2012b}. However, we estimate that the scintillation index for the full aperture is at least three orders of magnitude smaller than that of the 1.1~cm  square apertures, rendering this effect negligible in our final results.

\begin{figure}[t!]
\centering
\includegraphics[width=15cm]{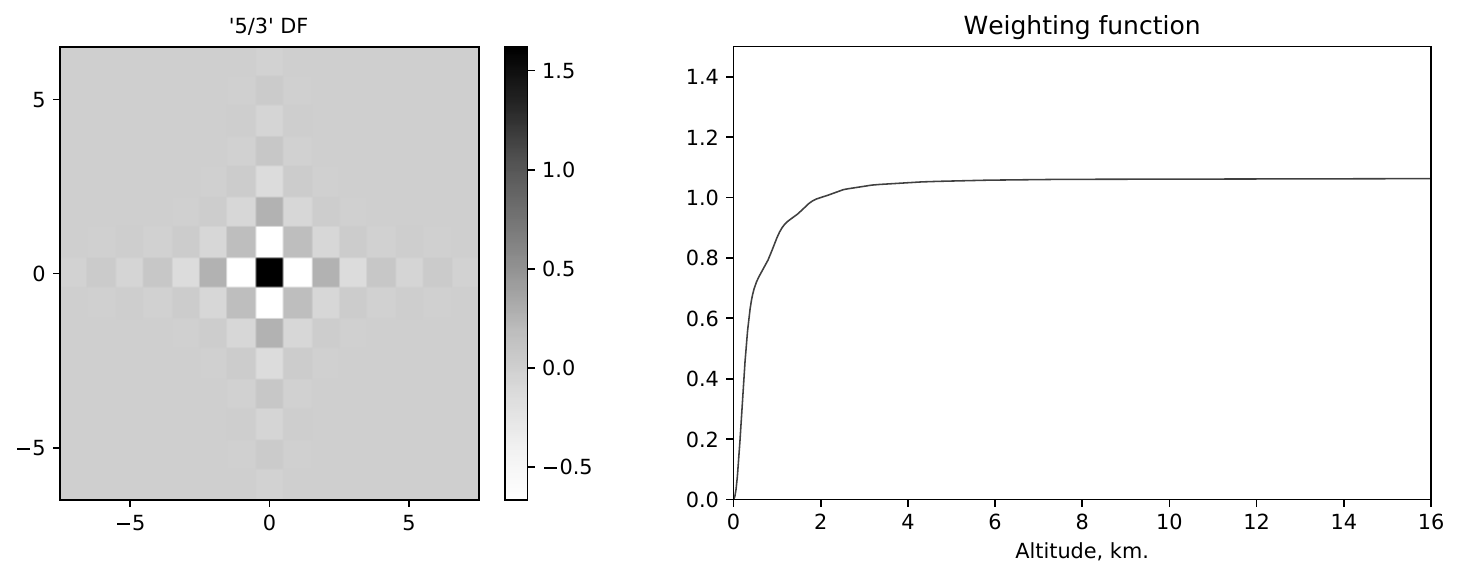}
\caption{Left panel: examples for digital filter impulse $\omega_{jk}$
for `$5/3$' digital filter.
In order to choose impulse normalization we assumed $11.68~\mathrm{mm}$ square apertures and spectral filter~$E(u)$ from~Fig.~\ref{fig:spectral_filter}.
Right panel: The weighting function for `$5/3$' digital filter corresponding to the case of $11.68~\mathrm{mm}$ square apertures tiling the $2.5~\mathrm{m}$ aperture.
Spectral filter~$E(u)$ is the same as shown at~Fig.~\ref{fig:spectral_filter}.
}
\label{fig:impulses}
\end{figure}

The mean flux has to be estimated in order to evaluate relative flux fluctuations as defined in (\ref{eq:relative_flux_fluctuation}). This mean flux varies spatially across the frame due to non-uniformities in the optical system and CCD detector response. To determine this spatial dependence we averaged the normalized data cube along the time axis to estimate mean pupil image representing the long exposure case. We found that a $1$~second average window provides optimal results, while longer windows (e.g. $1$~minute) introduce systematic biases in the auto-covariance estimates.  These biases predominantly affect pixels near the pupil edges and those obscured by dust particles in the optical path.

The final auto-covariance estimates are computed according to the Wiener-Khinchin theorem. First, two-dimensional Fourier transform is applied to the relative flux fluctuation data cube. Next, we select all frame pairs separated by time delay $t$, perform pixel-wise multiplication, and average the results. We use the following simple grid for time delay: $0,$ $1,$ $2,$ $4,$ $8$~frames, given frame rate is $100$ frames per second, this leads to $t = 0~\mathrm{ms},$ $10~\mathrm{ms},$ $20~\mathrm{ms},$ $40~\mathrm{ms},$ $80~\mathrm{ms}$. Finally, we apply the inverse transform to the averaged power spectra. Within our production core we use state of the art Fast Fourier Transform library~{\tt FFTW3}~\citep{FFTW05}.

The final result are estimates of $\gamma^{(t)}_{jk}$ as defined in Equation~(\ref{eq:wind_sigma_gamma}). Note that the pupil transfer coefficients $c_{jk}$~(equivalently represented by the pupil mask $\mathbf{I}_{0^\complement}(\rho_{nm})$) must also be preserved for subsequent analysis, as they appear explicitly at the right-hand side of equation (\ref{eq:wind_sigma_gamma}). The example of measurement is displayed in Fig.~\ref{fig:ACF_example}. One can see peaks corresponding to the turbulent layers with high wind speed, gradually departing from the origin as the delay value increases. These peaks are also wider, indicating high altitude. Narrow peak at the origin, visible only at conjugation altitude of $-2$~km is due to dome OT.






\section{Dome turbulence estimation}
\label{sec:estimation}


While in principle the estimated auto covariance functions can be used to recover both atmospheric OT and wind profiles \citep{Habib2006}, in this study we focus exclusively on estimating dome OT. We propose the following technique to estimate the latter. 

The dome OT estimation requires using auto-covariance functions with time delay, which enables spatial separation between dome OT and atmospheric OT. We employ the synthetic scintillation index from Equation~(\ref{eq:gamma_sum}) with $\Omega(\vb u) \sim u^{5/3}$. The corresponding impulse $\omega_{ij}$ from Equation~(\ref{eq:digital_filter}) is evaluated numerically, accounting for both the pupil transfer coefficients $c_{jk}$ and the square pixel aperture filter. The impulse is provided in Fig.~\ref{fig:impulses}, left panel. Note that $\omega_{jk}$ becomes negligible for sufficiently large indices $|j|,|k| \ge 10$, meaning the filter automatically excludes atmospheric OT when applied to time-delayed auto-covariance functions. The associated weighting function appears in Fig.~\ref{fig:impulses}, right panel, showing near-constant values for altitudes above $2~\mathrm{km.}$

In order to estimate dome OT $J_\mathrm{dome}$, the `$5/3$' digital filter is applied to auto covariance functions with all available time delays obtained for the virtual pupil:
\begin{equation}
J_\mathrm{dome} = \sum_{j,k} \omega_{jk} \gamma_{jk}, 
\end{equation}
where $\omega_{jk}$ is numerically evaluated impulse for `$5/3$' digital filter, normalized to provide OT power in correct units.

In principle, free atmosphere~(with respect to Fig.~{\ref{fig:impulses}}, right panel) OT power can be obtained by applying the filter to auto-covariance function without time delay for the real pupil. However, this would require reduction for the effect of finite exposure time, which is comparable to atmospheric time constant in our case. We leave this for the future work.


\subsection{Dome turbulence temporal evolution}
\label{subs:decay}

\begin{figure}[t!]
\centering
\includegraphics[width=9cm]{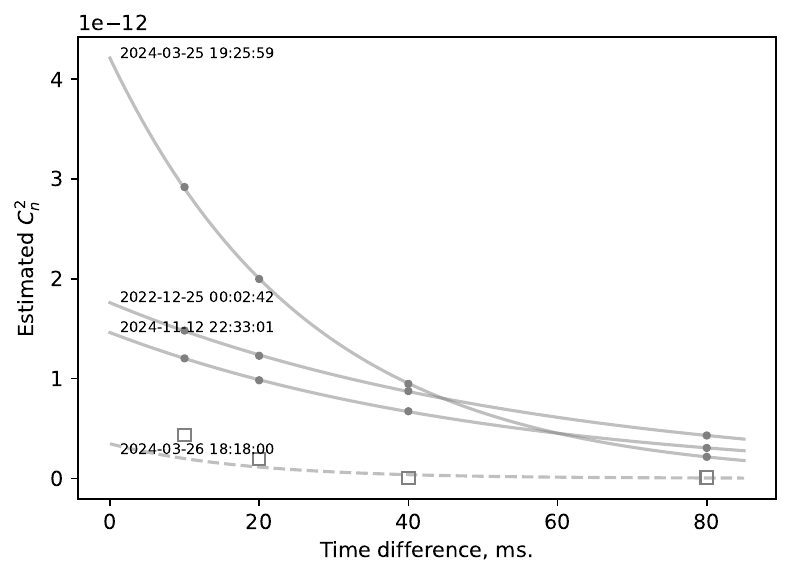}
\caption{
Dome OT power estimates for four representative measurements. Each measurement yield four different estimates corresponding to the available time delays (10, 20, 40, and 80~ms). Solid and dashed lines indicate the best-fit exponential models. Empty squares and dashed line mark the case showing poor agreement with exponential decay behavior.
}
\label{fig:decay}
\end{figure}

Our processing generates four separate dome OT estimates per measurement, corresponding to the four time delays of $10~\mathrm{ms},$ $20~\mathrm{ms},$ $40~\mathrm{ms},$ $80~\mathrm{ms}$. While Taylor's frozen flow hypothesis would predict consistent estimates across these delays, the observed dome OT decreases with time delay. This effect was initially identified by \citep{Osborn2023} based on data from two time delays, which they interpreted as following a linear relationship. Our dataset comprising multiple time delays demonstrates that the dome OT follows an exponential decay law with striking precision, as shown in Figure~\ref{fig:decay}.

The exponential fits reveal characteristic half-life times varying between $0.01$~s and $0.1$~s across different measurements. For the characterization of the fit quality we consider a sum of squared differences between logarithms of power and exponential fit value. We assume the exponential fit as successful if this value is less than 0.1. Cases where the exponential model shows poor agreement typically result from interference between atmospheric OT autocorrelation peaks and dome OT signals, occurring when wind speeds are particularly low, the example of such situation is provided in Fig.~\ref{fig:decay} (empty squares). 
Among the 88 measurements analyzed, 71 conform well to the exponential model, while the remainder would require complete characterization of atmospheric OT structure for proper interpretation \citep{Habib2006}.

The exponential decay behavior defies explanation through wind speed variance models, as Equation~(\ref{eq:wind_sigma_decay}), derived in Appendix~\ref{app:wind2}, unambiguously predicts power-law decay in such cases. No alternative explanation emerges within the standard Taylor hypothesis framework. \citep{Doob1942} demonstrated that normally distributed, continuous-time Markov processes necessarily exhibit exponential auto-covariance functions. Given that the laws of hydrodynamics are described by differential equations, it seems that velocity field temporal evolution would be indeed Markovian \citep{Nickelsen2014}.


\subsection{Comparison with SPP: outer scale}
\label{subs:outerscale}

\begin{figure}[t!]
\centering
\includegraphics[width=13cm]{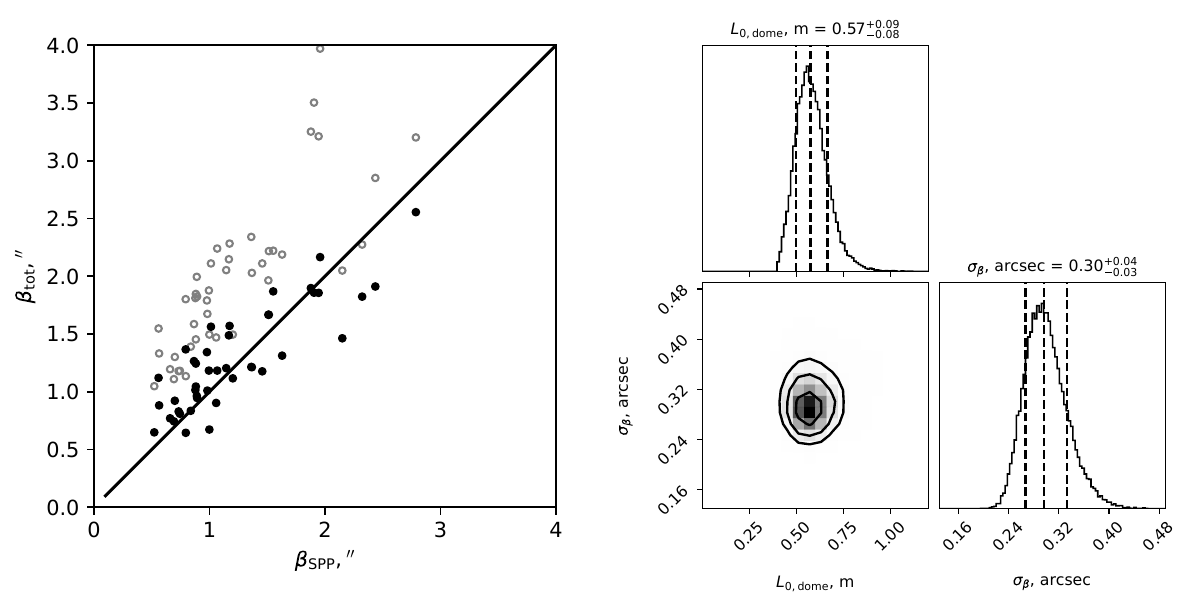}
\caption{
Left panel: DIQ predicted from dome OT power measured with Domecam and atmosphere OT power measured with MASS-DIMM. Empty circles stand for infinite outer scales, filled circles correspond to $L_{0,\mathrm{atm}}=25$~m and $L_{0,\mathrm{dome}}=$\outerscale~m.
Right panel: corner plot showing the posterior posterior probability distribution of parameters $L_{0,\mathrm{dome}}$ and $\sigma_\beta$, Equation (\ref{eq:logP_SPP_DC}).
}
\label{fig:SPP_comparison}
\end{figure}

\begin{figure*}[t]
\centering
\includegraphics[width=16cm]{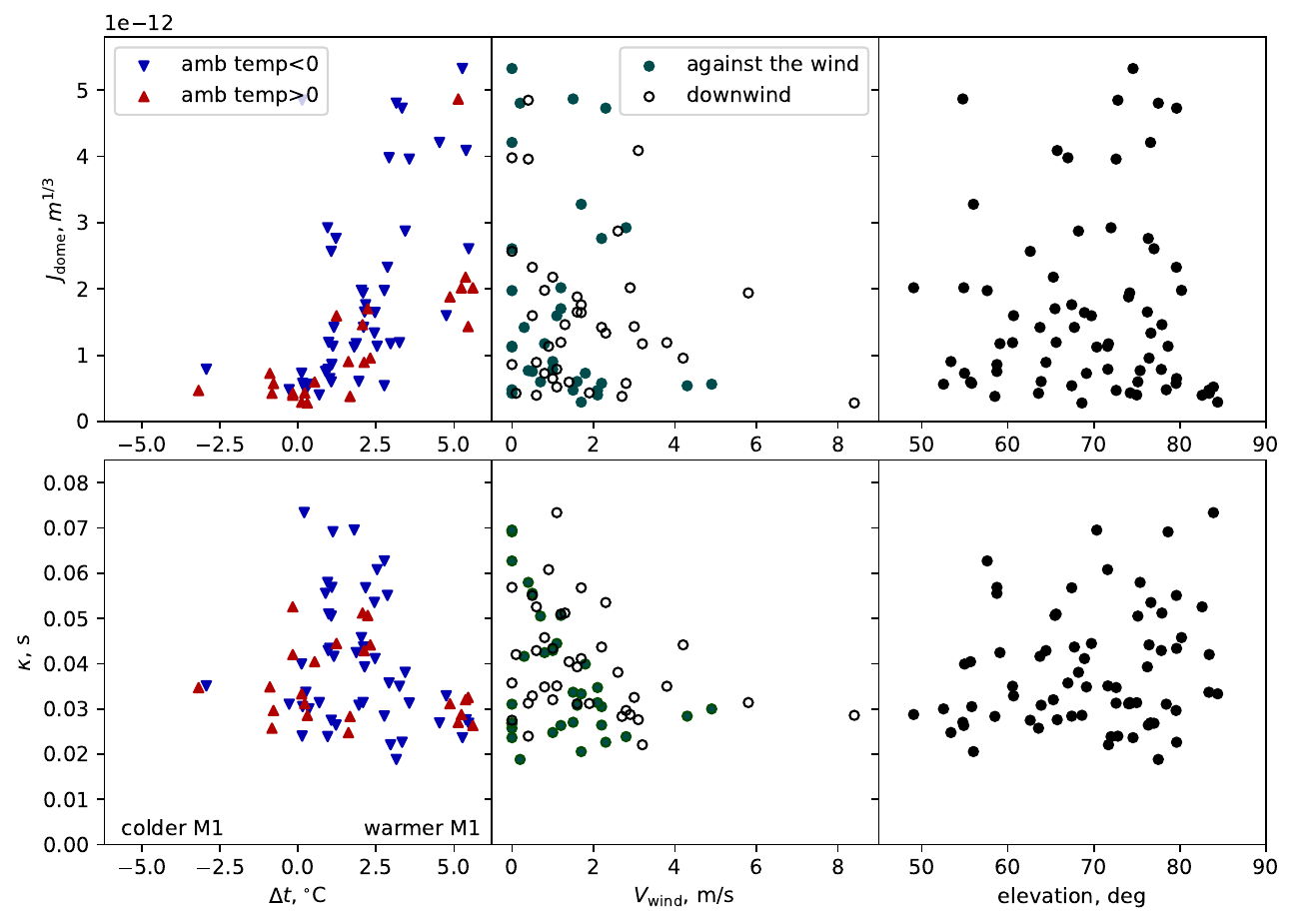}  
\caption{
Dependence of dome turbulence characteristics on environmental conditions. Top: Turbulence power ($J_\mathrm{dome}$) versus (left) temperature difference between M1 mirror and dome air ($\Delta t$), (center) wind speed ($V_\mathrm{wind}$), and telescope elevation. Bottom: Corresponding turbulence decay timescale ($\kappa_\mathrm{dome}$) relationships.
}
\label{fig:dome_amb}
\end{figure*}

Estimating delivered image quality (DIQ) at the telescope from turbulence integral measurements requires proper consideration of the turbulence outer scale $L_0$. While the free atmosphere typically exhibits $L_{0,\mathrm{atm}}\approx25$~m, for the dome OT $L_{0,\mathrm{dome}}$ is likely to be significantly smaller. 

As we will see soon, $L_{0,\mathrm{dome}}$ can be become even comparable to Fried radius. Due to this and also the fact that we want different outer scales for atmospheric and dome turbulence, we cannot employ approximation by A.A. Tokovinin \citep{Tokovinin2003} for the effect of outer scale on DIQ. Instead we computed it numerically. For a given power $J_\mathrm{dome}$ and outer scale $L_{0,\mathrm{dome}}$ we calculated structure function $\mathcal{D}_\phi$. The same was done for atmospheric OT, for $J_\mathrm{atm}$ we took results of simultaneous measurements with MASS--DIMM, scaled for the line of sight and wavelength of observations with Domecam. $L_{0,\mathrm{atm}}$ was fixed at 25~m.

Due to statistical independence of realizations of phase perturbations in dome OT and atmosphere OT, the total structure function of the resulting phase perturbations is a simple sum of structure functions of dome and atmosphere inputs. The optical transfer function of the system can be calculated as $T(f)=T_0(f)\exp\{-\mathcal{D}_\phi(\lambda f)/2\}$, where $T_0(f)$ is the diffraction limited optical transfer function \citep{Roddier1981}. Finally the DIQ $\beta_{tot}$ was computed as a FWHM of the corresponding point spread function.

The dome turbulence outer scale $L_{0, \mathrm{dome}}$ estimation relies on comparison between total seeing derived from Domecam and MASS--DIMM measurements $\beta_\mathrm{tot}$ and simultaneous SPP measurements $\beta_\mathrm{SPP}$. Some Domecam measurements have simultaneous series containing 1500 frames with 30 ms exposure of the same object obtained with the SPP. Processing of SPP data first subtracts bias frames, then averages images into 5-second intervals without recentering. We calculate FWHM for these 5-second effective exposure images, then average results across the full series. Out of 71 reliable estimations of $J_\mathrm{dome}$, 43 have simultaneous measurements of DIQ with SPP.

\begin{figure*}[t]
\centering
\includegraphics[width=16cm]{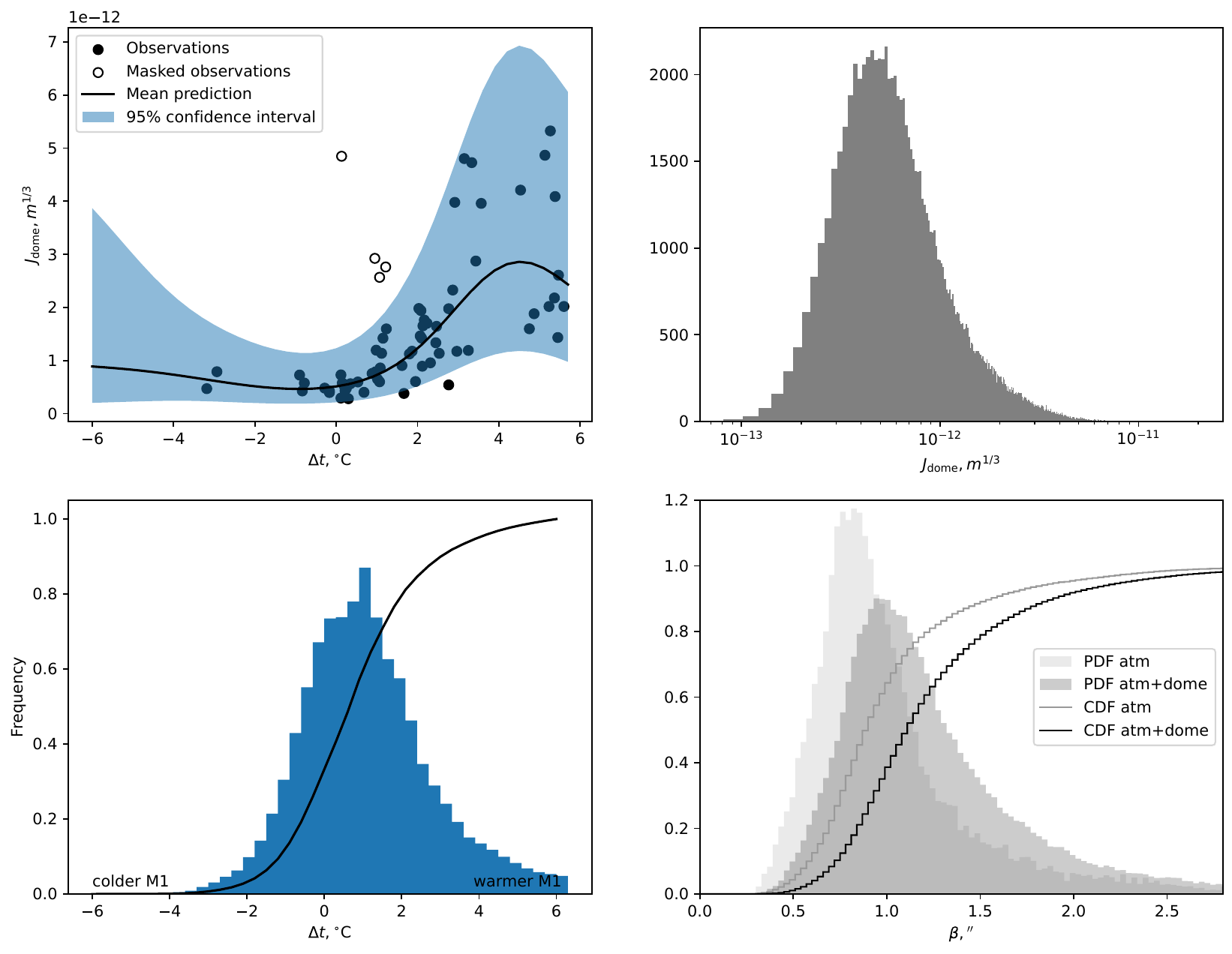}
\caption{
Upper left: gaussian regression fit of dependence of dome OT power on temperature difference $\Delta t$ between air inside the dome and M1 mirror. Lower left: histogram and CDF of $\Delta t$. Upper right: distribution of expected $J_\mathrm{dome}$. Lower right: distribution of expected DIQ with and without dome OT input.  
}
\label{fig:dome_prediction}
\end{figure*}

Figure~\ref{fig:SPP_comparison}, left panel compares the measured SPP seeing $\beta_\mathrm{SPP}$ with Domecam's predicted total seeing $\beta_\mathrm{tot}$ under assumption of infinite outer scales. While the data show strong correlation between $\beta_\mathrm{SPP}$ and $\beta_\mathrm{tot}$, a systematic tendency emerges where SPP measurements are typically smaller than Domecam predictions. A significant scatter in the data points may partially originate from telescope vibrations \citep{Strakhov2023}, though other contributing factors remain unclear.

We employed Bayesian inference to constrain  $L_{0, \mathrm{dome}}$, defining the posterior probability as 
\begin{equation}
P(L_{0, \mathrm{dome}},\sigma_\beta) = \frac{1}{\sqrt{2 \pi N \sigma_\beta^2}} \exp\Biggl\{ {-\frac{1}{2}\sum_{i=1}^N\frac{\bigl(\beta_{\mathrm{tot},i}(L_{0,\mathrm{dome}})-\beta_{\mathrm{SPP},i}\bigr)^2}{\sigma_\beta^2}} \Biggr\},   
\label{eq:logP_SPP_DC}
\end{equation}
where $\sigma_\beta^2$ quantifies additional seeing measurement variance, $i$ denotes individual observations, and $N$ is the total observation count. The free atmosphere outer scale was fixed at $L_{0, \mathrm{atm}}=25$~m. $\beta_\mathrm{tot}$ was calculated numerically as described above in this subsection.

Priors included uniform distributions for $L_{0,\mathrm{dome}}$ ($0.1-100$~m) and $\sigma_\beta^2$ ($0-1^{\prime\prime}$). Through Markov Chain Monte-Carlo sampling (32 walkers, $10^4$ iterations each), we obtained the posterior distribution, see Fig.~\ref{fig:SPP_comparison}, right panel. The results indicate $L_{0,\mathrm{dome}}=$\outerscaleErr~m and $\sigma_\beta=0.30^{\prime\prime}\pm0.03^{\prime\prime}$. The DIQ expected for these parameters agrees well with SPP data, as can be seen from Fig.~\ref{fig:SPP_comparison}, left panel.



\section{Results}

Our analysis estimated dome OT using digital filter for different time delays. For each measurement series, we extrapolated the turbulence-delay dependence to zero delay using the exponential law described in Subsection~\ref{subs:decay} of Section~\ref{sec:estimation}. We present two key parameters: the zero-delay dome turbulence intensity $J_\mathrm{dome}$ (top row of Figure~\ref{fig:dome_amb}) and the exponential timescale $\kappa_\mathrm{dome}$ (bottom row).

The left column of Figure~\ref{fig:dome_amb} shows these parameters as functions of the temperature difference $\Delta t$ between the M1 mirror and dome air (measured 2.8 m above floor level near the southern and western walls). The minimum dome OT $J_\mathrm{dome}\approx3-5\times10^{-13}$~m$^{2/3}$ occur when $\Delta t\in[-1^{\circ},1^{\circ}]$. When the mirror becomes warmer than the dome air $\Delta t>1^{\circ}$, $J_\mathrm{dome}$ increases significantly with $\Delta t$, likely due to convection of air. At $\Delta t \approx 5^{\circ}$, the OT intensity reaches $5\times10^{-12}$~m$^{2/3}$ which is 10--15 times stronger than the minimum values. Despite this clear temperature dependence, substantial scatter persists at large positive $\Delta t$ values, whose origin remains unclear. It is noteworthy that this effect is more pronounced during winter season, see upper left panel of Fig.~\ref{fig:dome_amb}. Otherwise, overall dependence of $J_\mathrm{dome}(\Delta t)$ appears the same for summer and winter.

For negative $\Delta t$ values, our dataset contains insufficient measurements to establish definitive trends in $J_\mathrm{dome}$ behavior. Preliminary evidence suggests a potential increase in $J_\mathrm{dome}$ for $\Delta t<0^{\circ}$, though less pronounced than the steep rise observed for $\Delta t>0^{\circ}$. As shown in the top right panel of Figure~\ref{fig:dome_amb}, we find no clear correlation between dome OT intensity and external wind speed $V_\mathrm{wind}$. There is weak evidence that for wind speed larger than 1~m/s it is preferable to point the telescope downwind when $\Delta t>2^{\circ}$C.  

The turbulence timescale $\kappa_\mathrm{dome}$ exhibits dependence on both $\Delta t$ and $V_\mathrm{wind}$ (lower row of Figure~\ref{fig:dome_amb}). The $\kappa_\mathrm{dome}(V_\mathrm{wind})$ relationship shows a distinct upper boundary that decreases with increasing wind speed, consistent with expectations that stronger winds accelerate turbulence evolution near the mirror surface. This effect is more pronounced when the telescope is pointed against the wind, as expected. However, we emphasize that this faster timescale does not correspond to reduced turbulence intensity. When examining temperature effects, the data reveal consistently rapid turbulence ($\kappa_\mathrm{dome}<0.03$~s) during conditions of warm M1 mirror ($\Delta t>3^{\circ}$). This suggests that thermal gradients may dominate turbulence dynamics in this regime.

From the right column of Fig.~\ref{fig:dome_amb} one can see a tentative correlation between dome OT and telescope elevation, with lower elevations showing reduced $J_\mathrm{dome}$. This may result from decreased overlap between the convective air column above the mirror and the stellar light path at these elevations.

\begin{figure}[t!]
\centering
\begin{tabular}{c}
\includegraphics[width=7cm]{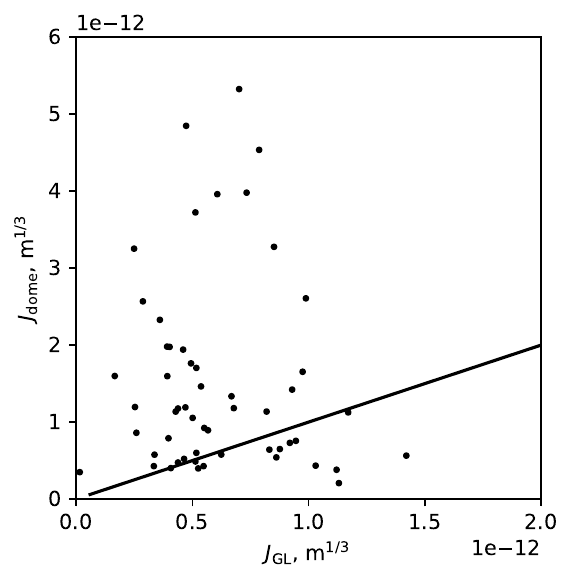} \\ 
\end{tabular}
\caption{
Comparison of $J_\mathrm{dome}$ measured by Domecam and OT power in ground layer measured by MASS-DIMM. Line marks equality of these quantities.}
\label{fig:GL_dome}
\end{figure}

It is also instructive to compare the OT power in the ground layer ($h<0.5$~km), as measured by MASS-DIMM to $J_\mathrm{dome}$, see Fig.~\ref{fig:GL_dome}. One can see that there is no correlation between those quantities, meaning that Domecam measurements is not biased by ground layer turbulence.

In order to quantify the effect of the dome OT on the images obtained in the 2.5-m telescope we performed a gaussian regression of dependence of $J_\mathrm{dome}$ on $\Delta t$, the result is displayed in the upper left panel of Fig.~\ref{fig:dome_prediction}. While the total standard deviation of $J_\mathrm{dome}$ is $1.26\times10^{-12}$~m$^{1/3}$, the standard deviation of residuals is $0.84\times10^{-12}$~m$^{1/3}$.

Lower left panel of Fig.~\ref{fig:dome_prediction} contains the distribution of $\Delta t$, observed during periods when the dome was open. For construction of this histogram we used 4050~h of data obtained from 2021-07-18 to 2024-04-21. The quartiles of this distribution are the following $-0.03^{\circ}$C, $0.98^{\circ}$C, $2.10^{\circ}$C.

The distribution of $\Delta t$ and gaussian fit of $J_\mathrm{dome}(\Delta t)$ were combined to estimate the distribution of $J_\mathrm{dome}$, see the upper right panel of Fig.~\ref{fig:dome_prediction}. The measurements of atmospheric OT power, secured with MASS-DIMM in period from 2022-11-08 to 2025-03-13, and dome OT power distribution were used to build distributions of DIQ, with and without the effect of dome OT, using numerical method described in previous subsection. For the outer scale $L_0$ we took \outerscale~m under the dome (section~\ref{subs:outerscale}) and usual 25~m for the free atmosphere OT. We assumed that the free atmosphere and dome OT are not correlated. These distributions are displayed in lower right panel of Fig.~\ref{fig:dome_prediction}. The median DIQ at 500~nm at zenith is
\medianDIQwithDome and \medianDIQwithoutDome, with and without dome OT, respectively. 

\section{Discussion and conclusion}

Optical turbulence (OT) inside the dome of the telescope significantly affects the performance of the latter. For characterization of dome OT we have built an instrument Domecam. The Domecam is conceptually similar to Single Star SCIDAR and Dome Turbulence Monitor \citep{Osborn2023}, it measures the fluctuations of intensity of radiation of a distant point-like source --- a star --- at the plane optically conjugated by 2 km below the entrance pupil. These measurements are used to compute autocorrelation with a temporal delay. In a resulting autocorrelation dome OT manifests itself as a turbulent layer at 0 km altitude and, more importantly, having zero wind speed. Turbulent layers located in the free atmosphere generally have wind speeds larger than $2-3$~m/s.

Here we consider several methods of estimation of dome OT power $J_\mathrm{dome}$ from autocorrelation of intensity fluctuations. The most robust results are obtained using the so-called ``digital filtering'' approach. In this method one computes the sum of a product of Fourier spectra of autocorrelation and spectral filter of a certain shape. We present measurements obtained with Domecam at the 2.5-m telescope of Caucasian Observatory of Sternberg Astronomical Institute Lomonosov MSU. Measurements were obtained in 2022-2024 in various ambient conditions. Simultaneous observations with the MASS-DIMM instrument located 20 m from the tower of the 2.5-m telescope showed that scintillation indexes directly measured with MASS-DIMM are predicted by Domecam with typical correlation coefficients of $0.95-0.99$.

Domecam is sensitive to refractive fluctuations having typical size of several cm. The question arises what is the quantitative effect of dome OT power estimated at these spatial scales on the image in a large telescope. Remind that FWHM of long exposure image - delivered image quality (DIQ) - depends on refractive index fluctuations at all spatial scales. Using simultaneous measurements of DIQ with the SPeckle Polarimeter instrument and Domecam measurements we have shown that Domecam can predict DIQ if one assumes von K\'{a}rm\'{a}n structure function of phase fluctuations with outer scale parameter $L_{0,\mathrm{dome}}=$\outerscaleErr~m for the dome OT. This value hints that in case of 2.5-m telescope dome OT is concentrated in a relatively thin convective layer above the main mirror.

This is further corroborated by the fact that the difference in temperature between the main mirror of the telescope and the air under dome $\Delta t$ has the the greatest effect on the power $J_\mathrm{dome}$. The latter starts to rise when $\Delta t$ as low as $1^{\circ}$. At the same time we obtained tentative evidence that for periods of negative $\Delta t$ the dependence $J_\mathrm{dome}$ on $|\Delta t|$ is not as steep. Using the distributions of $\Delta t$ and OT power in the free atmosphere, and outer scale value under the dome, we estimated expected distributions of DIQ for the 2.5-m telescope. Its median is \medianDIQwithDome at 500~nm, meanwhile in the absence of dome OT one would expect a median DIQ of \medianDIQwithoutDome.

We found out that dome OT decorrelates with characteristic time $\kappa_\mathrm{dome}=30-70$~ms. Exponential law of this decorrelation does not comply with assumption that dome OT is a combination of frozen turbulent layers with the normal distribution of wind speeds. Apparently there is significant violation of Taylor frozen flow hypothesis, which should be taken into account while modeling of dome OT influence, e.g. on adaptive optics. $\kappa_\mathrm{dome}$, characterizing temporal evolution of dome OT, decreases with the rise of $\Delta t$ and in stronger wind conditions. Quantitative analysis of factors affecting $\kappa_\mathrm{dome}$ will be conducted in a future work.

\vspace{0.5cm}

{\bf Data availability.} Zenodo repository at \href{https://zenodo.org/records/15622120}{https://zenodo.org/records/15622120} contains intermediate and final processing products: auto--covariances of scintillation patterns, estimations of dome OT power and time constant, accompanied by the data from the MASS-DIMM, SPP and weather station. Due to the large size of the raw data, it cannot be easily shared; however, it may be obtained from the authors upon reasonable request.

\vspace{0.5cm}

{\bf Code availability.} Control and processing software of Domecam can be found in mercurial repository at: \href{https://curl.sai.msu.ru/hg/home/matwey/domecam}{https://curl.sai.msu.ru/hg/home/matwey/domecam}.

\vspace{0.5cm}

{\bf Disclosures.}
The authors declare there are no financial interests, commercial affiliations, or other potential conflicts of interest that have influenced the objectivity of this research or the writing of this paper




\appendix

\section{The effects of finite exposure time and wind velocity dispersion within the turbulent layer}
\label{app:wind2}

For the purposes of this appendix, to simplify equations, we omit explicit ${\bf w}(z)$ assuming that the reader can either substitute $j$ with $j - w_x(z) t / \Delta$ or consider that $j$ is currently counted from $w_x(z) t / \Delta$ pixel.

\begin{figure}[b!]
\centering
\includegraphics[height=6cm]{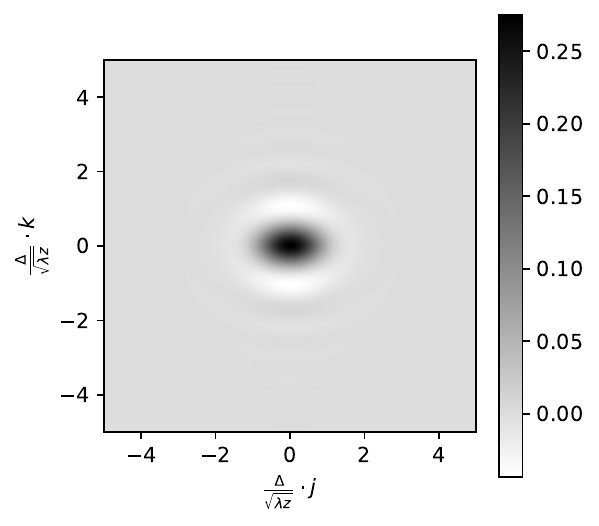}
\caption{
Autocorrelation function for relative flux fluctuation in case of infinitely small aperture and wind shear ${\tau w(z)} = {\Delta}$.
}
\label{fig:acf_point_wind}
\end{figure}

Building upon the approached developed in \citep{Tokovinin2002} and \citep{Kornilov2011}, we derive the following equation incorporating finite exposure $\tau$ effect:
\begin{multline}
    \label{eq:wind_shear_gamma}
    \gamma_{jk} = 9.69 \cdot 10^{-3} \cdot \frac{16 \pi^2}{\lambda_0^{\frac{7}{6}}} z^{\frac{5}{6}} C_n^2(z) dz c_{jk} \int \vb{d^2 u} \, {u}^{-\frac{11}{3}} E(u) A\left(\frac{D}{\sqrt{\lambda_0 z}}\vb{u}\right) W_{\tau}\left(\frac{\tau}{\sqrt{\lambda_0 z}} (\vb{w}(z), \vb{u})\right)\times \\ \times \exp\left(-2\pi i \frac{\Delta}{\sqrt{\lambda_0 z}} \left(j u_x + k u_y\right)\right),
\end{multline}
where $\vb{w}(z)$ is the wind velocity vector at altitude $z$, $(\cdot, \cdot)$ denotes inner product, and the wind shear filter $W_{\tau}(u) \equiv \sinc^2 \left( u \right).$ Comparison with equation~(\ref{eq:convolution_gamma}) reveals that Equation~(\ref{eq:wind_shear_gamma}) describes an anisotropic blurring along the wind direction. The blurring kernel is a narrow one-dimensional triangle function with maximum $\frac{\Delta}{\tau w(z)}$ and unit integrated area. Fig.~\ref{fig:acf_point_wind} illustrates this finite exposure effect for the case $\frac{\Delta}{\tau w(z)} = 1$.

Additionally, consider what happens if the turbulence layers is composed of multiple sublayers with distinct velocities. We assume that the wind velocity vector follows a normal distribution with mean $\vb w(z)$ and isotropic variance $\sigma^2_w(z)$ \citep{Habib2006}. An additional isotropic Gaussian blurring effect occurs when measuring the autocorrelation with time delay $t$. This blurring is characterized by the filter:
\begin{equation}
    \label{eq:wind_sigma_filter}
    W_{\sigma}( u) \equiv \exp\left(-u^2 \right).
\end{equation}
The autocorrelation function for relative flux fluctuations becomes:
\begin{multline}
    \label{eq:wind_sigma_gamma}
    \gamma^{(t)}_{jk} = 9.69 \cdot 10^{-3} \cdot \frac{16 \pi^2}{\lambda_0^{\frac{7}{6}}} z^{\frac{5}{6}} C_n^2(z) dz c_{jk} \int \vb{d^2 u} \, {u}^{-\frac{11}{3}} E(u) A\left(\frac{D}{\sqrt{\lambda_0 z}}\vb{u}\right) W_{\sigma}\left(\frac{2\pi \sigma^2_w(z) t}{\sqrt{\lambda_0 z}} u\right)\times \\ \times\exp\left(-2\pi i \frac{\Delta}{\sqrt{\lambda_0 z}} \left(j u_x + k u_y\right)\right).
\end{multline}

For an infinitely small aperture~($A({\vb u})=1$), equation~(\ref{eq:wind_sigma_gamma}) admits a closed-form solution, yielding the the following dependence of $\gamma_{00}$ on the time delay $t$: 
\begin{multline}
    \label{eq:wind_sigma_decay}
    \frac{\gamma^{(t)}_{00}}{\gamma^{(0)}_{00}} = \cos \left( \frac{7}{6}\,\arctan \left( \kappa \right)  \right)  \left( 1+{\kappa}^{2} \right) ^{-{{7}/{12}}} \left(1 - \kappa^2\right) 
    + 2\,\kappa\sin \left( \frac{7}{6}\,\arctan \left( \kappa \right)  \right)  \left( 1+{\kappa}^{2} \right) ^{-{{7}/{12}}}-\\
    -{\kappa}^{5/6}\sec \left( \frac{\pi}{12}  \right) \sin \left( \frac{5}{12}\,\pi  \right) \csc \left( \frac{\pi}{12} \right) + \\
 +{{\csc \left( \frac{\pi}{12}  \right) \sin \left( \frac{5}{12}\,\pi  \right) }{
 \left( 1+{\kappa}^{2} \right) ^{-1/12}}} \times \left(\kappa \cos \left( \frac{1}{6}\,\arctan \left( \kappa \right)
 \right) - \sin \left( \frac{1}{6}\,\arctan \left( \kappa \right)  \right) \right),
\end{multline}
where $\kappa \equiv {2\pi\sigma^4_w t^2} / {(\lambda z)}$. The asymptotic behavior ${\gamma^{(t)}_{00}} / {\gamma^{(0)}_{00}} \sim \kappa^{-6/5}$ emerges as $\kappa \rightarrow +\infty$. 

\section{Comparison with MASS}
\label{sec:mass}

\begin{figure}[b!]
\centering
\includegraphics[width=6cm]{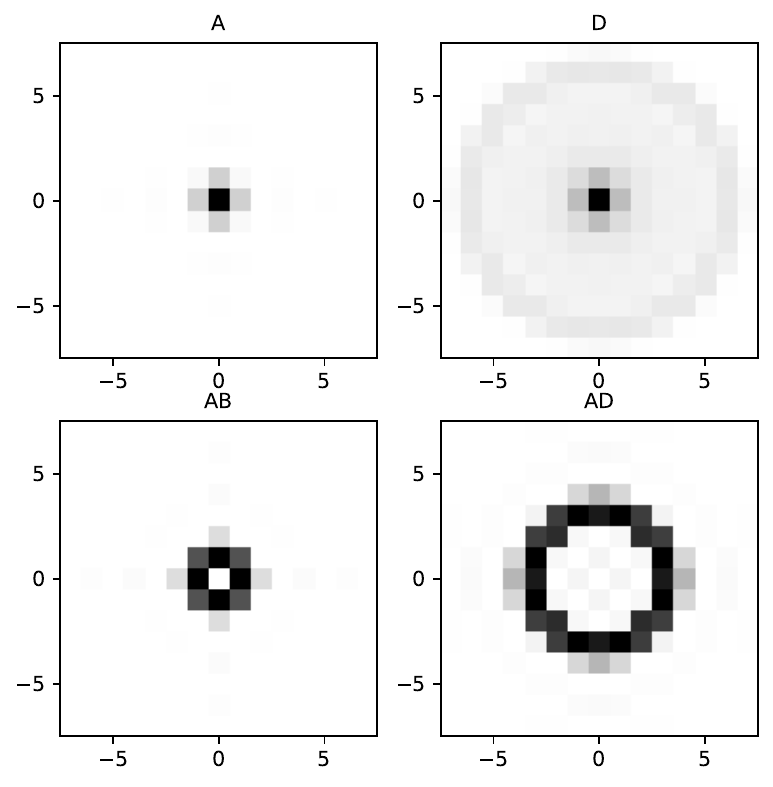}
\caption{Examples for digital filter impulse $\omega_{jk}$
for MASS scintillation indices in apertures A and D (upper pair)
and MASS covariance between apertures A and B, and A and D (lower pair).
Pixel density represents weight magnitude $\omega_{jk}$, while white color denotes zero.
Central pixel of each picture corresponds to $\omega_{00}$.
}
\label{fig:mass_weights}
\end{figure}

\begin{figure}[b!]
\centering
\includegraphics[clip=true, trim={0cm 4cm 0cm 4cm},width=16cm]{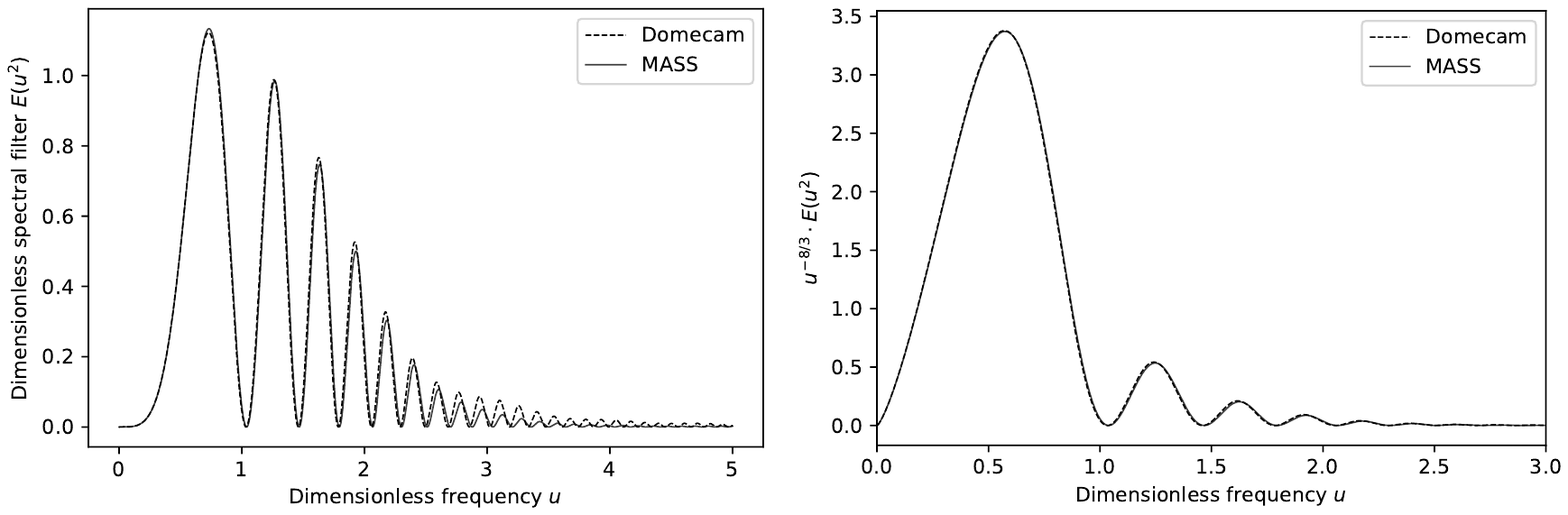} 
\caption{
Comparison of MASS and Domecam spectral filters $E(u)$.
Left panel: dimensionless spectral filter~$E(u)$ is shown.
Right panel: the same with respect to angle-averaged scintillation power spectra, i.e. $u^{-\frac{8}{3}} E(u)$.
Solid line is the MASS spectral filter~($\lambda_{0} = 516\,\mathrm{nm}$),
dashed line is the Domecam spectral filter~($\lambda_{0} = 742\,\mathrm{nm}$).}
\label{fig:spectral_filter}
\end{figure}

\begin{table}[b!]
\begin{center}
\begin{tabular}{ | l | c | c | }
\hline
& Inner size & Outer size \\
\hline
\hline
Domecam pixel size & \multicolumn{2}{| c |}{$11.68\,\mathrm{mm}$} \\
\hline
\hline
MASS~A & $0\,\mathrm{mm}$ & $20.57\,\mathrm{mm}$ \\
MASS~B & $21.06\,\mathrm{mm}$ & $34.83\,\mathrm{mm}$ \\
MASS~C & $35.64\,\mathrm{mm}$ & $62.37\,\mathrm{mm}$ \\
MASS~D & $63.18\,\mathrm{mm}$ & $89.10\,\mathrm{mm}$ \\
\hline
\end{tabular}
\end{center}
\caption{Comparison between geometrical scales of Domecam and MASS instrument attached to RCX400 telescope.}
\label{table:mass_geometry}
\end{table}

\begin{figure*}[t!]
\centering
\includegraphics[width=16cm]{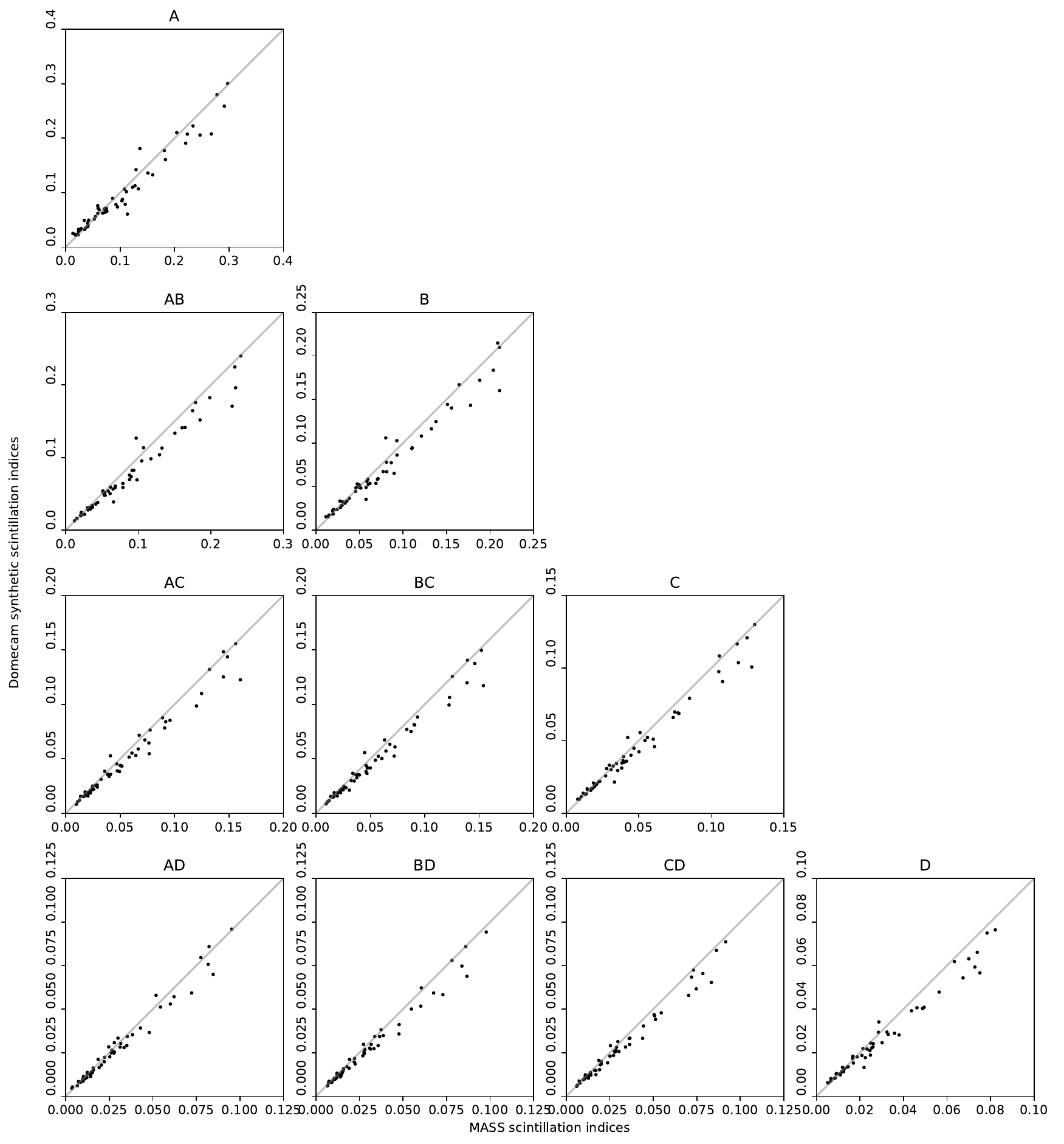}
\caption{Comparison of scintillation indices between direct MASS measurements and Domecam-synthesized values.
Labels $A,$ $B,$ $C,$ and $D$ represent scintillation indices in individual apertures, while
$AB,$ $AC,$ etc. indicate covariances between aperture pairs. 
Individual measurements are shown by black points.
Gray lines mark the perfect agreement.
}
\label{fig:mass_indices}
\end{figure*}

In order to evaluate Domecam instrument design and alignment on the telescope, we performed a comparison between Domecam and MASS instrument installed on the RCX400 telescope as part of the automatic seeing monitor of the observatory. The MASS instrument, featuring $4$ different apertures, produces $4$ scintillation indices and $6$ inter-aperture covariances. These quantities are usually used for free atmosphere OT profile restoration. Employing the digital filtering technique described in equation~(\ref{eq:gamma_sum}), we independently obtained the same quantities from the Domecam's ACF, effectively simulating the four-aperture configuration of MASS. For each MASS scintillation index and covariance we computed the corresponding digital filter $\Omega(\bf f)$ and its digital filter impulse $\omega_{jk}$. Fig.~\ref{fig:mass_weights} illustrates impulse patterns for two MASS scintillation indices and two MASS covariances. Synthetic MASS indices, generated from Domecam data, enabled direct performance comparison between Domecam and MASS.

This comparison has advantages over comparing OT profiles from both instruments. First, it avoids solving computationally and methodologically challenging inverse problems. Second, note that synthetic indices are independent of OT model assumptions, isolating the instruments' flux fluctuation measurement capabilities. OT profile recovery algorithms are out of the scope of current test.

This comparative approach presents several inherent limitations that require careful consideration. First, the exposure times must be precisely matched between instruments, as any discrepancy would lead to divergent wind shear effects and consequently inconsistent scintillation indices. Second, OT distribution in the atmosphere can potentially be nonisotropic and the same object must be measured with both instruments. And even if the distribution is isotropic the wind speed projection will be the same only for the same direction. Third, Domecam and MASS have different spectral response curves. Corresponding spectral filters for both instruments are given in Fig.~\ref{fig:spectral_filter}. While dimensionless spectral responses $E(u)$ are almost the same, the instruments have substantially different equivalent wavelength~$\lambda_{0}$. Fourth, some fundamental mathematical limits of synthetic indices approximation are examined in Section~\ref{sec:acf}.

Although these limitations cannot be completely eliminated, we implemented several measures to minimize their impact. To address potential issues arising from atmospheric anisotropy and wind speed projection differences, all observations were conducted simultaneously on identical targets using both Domecam and MASS instruments. While the standard MASS exposure time is $1\,\mathrm{ms}$, we utilized $4\,\mathrm{ms}$ exposures for this comparison. These longer exposures were digitally computed from raw MASS data to facilitate wind profile reconstruction, following the methodology described in \citep{Kornilov2012a}. The digital filter technique provides a valid approximation for MASS aperture filters despite Domecam's smaller pixel size relative to the smallest MASS aperture (see Table~\ref{table:mass_geometry} for detailed specifications). Additionally, all MASS scintillation indices were corrected for Poisson noise, scattered light, and PMT nonlinearity using the standard {\tt atmos} processing software.


The spectral difference between instruments presents the most significant challenge. Nevertheless it can be addressed using good agreement in the dimensionless spectral responses $E(u)$ and spectral-geometry scintillation similarity revealed by applying dimensionless frequencies $\vb u$. By comparing Equations~(\ref{eq:scintillation_index}) with~(\ref{eq:gamma_sum}) under the assumptions that $E(u)$ is identical for both instruments and $D = \Delta$, we derive the required digital filter form:
\begin{equation}
\Omega(\vb u) = \left(\frac{\lambda_{0}}{\lambda'_{0}}\right)^{\frac{7}{6}} \cdot \frac{A'\left(\frac{D'}{D} \sqrt{\frac{\lambda_{0}}{\lambda'_{0}}} \vb u\right)}{A\left(\vb u\right)},
\end{equation}
where $A'$ is the aperture filter for the considered MASS aperture evaluated using Equation~(\ref{eq:af_annular}), $D'$ is the MASS aperture scale chosen from Table~\ref{table:mass_geometry}, and $\lambda'_{0}$ is the MASS equivalent wavelength. While the equivalent wavelengths depend on the measured star spectral class for both instruments, it is remarkable that the value $\sqrt{{\lambda_{0}} / {\lambda'_{0}}} \approx 1.2$ is quite stable over different source spectral energy distributions.

Additionally, to account for the wind shear, the Domecam exposure $\tau$ has to be chosen as $\tau' \sqrt{{\lambda_{0}} / {\lambda'_{0}}}$ where $\tau'$ denotes the MASS exposure. For this purpose, while the MASS exposure $\tau' = 4\,\mathrm{ms}$, the Domecam exposure $\tau = 5\,\mathrm{ms}$ which is $1.2$ times longer.

Fig.~\ref{fig:mass_indices} provides a comparison between directly measured MASS indices and their Domecam-derived synthetic counterparts. The dataset consists of $23$ individual simultaneous measurements carried out from December~2022 through March~2024. The indices are not only highly correlated (Pearson correlation coefficient is between $0.95$ and $0.99$ for different indices) with relatively low spread, but their values are clearly in a good agreement.

\acknowledgments 

Unfortunately, one of co-authors of this work, V. G. Kornilov, passed away in 2021. For decades, he supervised optical turbulence studies at the Sternberg Astronomical Institute. We are indebted to him for his insights and guidance. This particular work was initiated by V. G. Kornilov; he developed the instrument's hardware and the analysis framework. The fact that he is the first author underscores his leading role in this study.

We are grateful to A.A. Tokovinin for sharing his insights about dome seeing measurements and valuable comments on the manuscript. We also thank the staff of the CMO SAI MSU for their assistance with the measurements. The study was conducted under the state assignment of Lomonosov Moscow State University. We acknowledge the comments by the anonymous referees, which helped us improve the presentation. LLM Deep Seek was used for grammar clean-up.

\bibliography{references} 

\begin{thebibliography}{10}

\bibitem{Bowen1964}
{Bowen}, I.~S., ``{Telescopes},'' {\em \aj}~{\bf 69},  816 (Dec. 1964).

\bibitem{Kornilov2012}
{Kornilov}, V., {Sarazin}, M., {Tokovinin}, A., {Travouillon}, T., and {Voziakova}, O., ``{Comparison of the scintillation noise above different observatories measured with MASS instruments},'' {\em Astronomy and Astrophysics}~{\bf 546},  A41 (Oct. 2012).

\bibitem{Kornilov1979}
{Kornilov}, V.~G., ``{Investigation of Optical Inhomogeneities of the Surface Layer of the Atmosphere by the Acoustic Method},'' {\em \sovast}~{\bf 23},  623 (Oct. 1979).

\bibitem{Bustos2018}
{Bustos}, E. and {Tokovinin}, A., ``{Dome seeing monitor and its results for the 4m Blanco telescope},'' in [{\em Ground-based and Airborne Telescopes VII}{\nolinebreak\hspace{0.1em}]},  {Marshall}, H.~K. and {Spyromilio}, J., eds., {\em Society of Photo-Optical Instrumentation Engineers (SPIE) Conference Series} {\bf 10700},  107000Q (July 2018).

\bibitem{Munro2023}
{Munro}, J., {Hansen}, J., {Travouillon}, T., {Grosse}, D., and {Tokovinin}, A., ``{Dome seeing analysis of the Anglo-Australian Telescope},'' {\em Journal of Astronomical Telescopes, Instruments, and Systems}~{\bf 9},  017004 (Jan. 2023).

\bibitem{Vernin1973}
{Vernin}, J. and {Roddier}, F., ``{Experimental determination of two-dimensional spatiotemporal power spectra of stellar light scintillation. Evidence for a multilayer structure of the air turbulence in the upper troposphere.},'' {\em Journal of the Optical Society of America (1917-1983)}~{\bf 63},  270--273 (Jan. 1973).

\bibitem{Avila1997}
{Avila}, R., {Vernin}, J., and {Masciadri}, E., ``{Whole atmospheric-turbulence profiling with generalized scidar},'' {\em \ao}~{\bf 36},  7898--7905 (Oct. 1997).

\bibitem{Klueckers1998}
{Klueckers}, V.~A., {Wooder}, N.~J., {Nicholls}, T.~W., {Adcock}, M.~J., {Munro}, I., and {Dainty}, J.~C., ``{Profiling of atmospheric turbulence strength and velocity using a generalised SCIDAR technique},'' {\em \aaps}~{\bf 130},  141--155 (May 1998).

\bibitem{Avila2000}
{Avila}, R., {Vernin}, J., {Chun}, M.~R., and {Sanchez}, L.~J., ``{Turbulence and wind profiling with generalized scidar at Cerro Pachon},'' in [{\em Adaptive Optical Systems Technology}{\nolinebreak\hspace{0.1em}]},  {Wizinowich}, P.~L., ed., {\em Society of Photo-Optical Instrumentation Engineers (SPIE) Conference Series} {\bf 4007},  721--732 (July 2000).

\bibitem{Habib2006}
{Habib}, A., {Vernin}, J., {Benkhaldoun}, Z., and {Lanteri}, H., ``{Single star scidar: atmospheric parameters profiling using the simulated annealing algorithm},'' {\em Monthly Notices of the Royal Astronomical Society}~{\bf 368},  1456--1462 (May 2006).

\bibitem{Errazzouki2024}
{Errazzouki}, Y., {Habib}, A., {Jabiri}, A., {Sabil}, M., {Benkhaldoun}, Z., {Azhari}, Y.~E., {Azagrouze}, O., and {Chafi}, J., ``{Single star SCIDAR: Atmospheric parameters profiling using the power spectrum of scintillation},'' {\em Astronomy and Computing}~{\bf 47},  100817 (Apr. 2024).

\bibitem{Osborn2023}
{Osborn}, J. and {Alaluf}, D., ``{Line-of-sight optical Dome Turbulence Monitor},'' {\em \mnras}~{\bf 525},  1936--1940 (Oct. 2023).

\bibitem{Shatsky2020}
{Shatsky}, N., {Belinski}, A., {Dodin}, A., {Zheltoukhov}, S., {Kornilov}, V., {Postnov}, K., {Potanin}, S., {Safonov}, B., {Tatarnikov}, A., and {Cherepashchuk}, A., ``{The Caucasian Mountain Observatory of the Sternberg Astronomical Institute: First Six Years of Operation},'' in [{\em Ground-Based Astronomy in Russia. 21st Century}{\nolinebreak\hspace{0.1em}]},  {Romanyuk}, I.~I., {Yakunin}, I.~A., {Valeev}, A.~F., and {Kudryavtsev}, D.~O., eds.,  127--132 (Dec. 2020).

\bibitem{Kornilov2007}
{Kornilov}, V., {Tokovinin}, A., {Shatsky}, N., {Voziakova}, O., {Potanin}, S., and {Safonov}, B., ``{Combined MASS-DIMM instruments for atmospheric turbulence studies},'' {\em Monthly Notices of the Royal Astronomical Society}~{\bf 382},  1268--1278 (Dec. 2007).

\bibitem{Potanin2017}
{Potanin}, S.~A., {Gorbunov}, I.~A., {Dodin}, A.~V., {Savvin}, A.~D., {Safonov}, B.~S., and {Shatsky}, N.~I., ``{Analysis of the optics of the 2.5-m telescope of the Sternberg Astronomical Institute},'' {\em Astronomy Reports}~{\bf 61},  715--725 (Aug. 2017).

\bibitem{Dodin2019}
{Dodin}, A., {Grankin}, K., {Lamzin}, S., {Nadjip}, A., {Safonov}, B., {Shakhovskoi}, D., {Shenavrin}, V., {Tatarnikov}, A., and {Vozyakova}, O., ``{Analysis of colour and polarimetric variability of RW Aur A in 2010-2018},'' {\em Monthly Notices of the Royal Astronomical Society}~{\bf 482},  5524--5541 (Feb. 2019).

\bibitem{Kornilov2021}
{Kornilov}, V., {Safonov}, B., and {Kornilov}, M., ``{Useful relations for the analysis of stellar scintillation at the entrance pupil of a telescope},'' {\em Journal of the Optical Society of America A}~{\bf 38},  1284 (Sept. 2021).

\bibitem{mpmath}
mpmath~development team, T., {\em mpmath: a {P}ython library for arbitrary-precision floating-point arithmetic (version 1.3.0)} (2023).
\newblock {\tt http://mpmath.org/}.

\bibitem{Tokovinin2003}
Tokovinin, A.~A., ``Polychromatic scintillation,'' {\em Journal of the Optical Society of America A}~{\bf 20},  686--689 (Apr 2003).

\bibitem{Kornilov2020}
{Kornilov}, M.~V., ``{Maximum Likelihood Estimation for Disk Image Parameters},'' {\em IEEE Signal Processing Letters}~{\bf 27},  1480--1484 (Jan. 2020).

\bibitem{Strakhov2023}
{Strakhov}, I.~A., {Safonov}, B.~S., and {Cheryasov}, D.~V., ``{Speckle Interferometry with CMOS Detector},'' {\em Astrophysical Bulletin}~{\bf 78},  234--258 (June 2023).

\bibitem{Kornilov2010}
{Kornilov}, V., {Shatsky}, N., {Voziakova}, O., {Safonov}, B., {Potanin}, S., and {Kornilov}, M., ``{First results of a site-testing programme at Mount Shatdzhatmaz during 2007-2009},'' {\em Monthly Notices of the Royal Astronomical Society}~{\bf 408},  1233--1248 (Oct. 2010).

\bibitem{Otsu1979}
Otsu, N., ``A threshold selection method from gray-level histograms,'' {\em IEEE Transactions on Systems, Man, and Cybernetics}~{\bf 9}(1),  62--66 (1979).

\bibitem{Kornilov2012b}
Kornilov, V., ``Stellar scintillation on large and extremely large telescopes,'' {\em Monthly Notices of the Royal Astronomical Society}~{\bf 426},  647--655 (10 2012).

\bibitem{FFTW05}
Frigo, M. and Johnson, S.~G., ``The design and implementation of {FFTW3},'' {\em Proceedings of the IEEE}~{\bf 93}(2),  216--231 (2005).
\newblock Special issue on ``Program Generation, Optimization, and Platform Adaptation''.

\bibitem{Doob1942}
Doob, J.~L., ``The brownian movement and stochastic equations,'' {\em Annals of Mathematics}~{\bf 43}(2),  351--369 (1942).

\bibitem{Nickelsen2014}
Nickelsen, {\em Markov Processes linking Thermodynamics and Turbulence}, PhD thesis (08 2014).

\bibitem{Roddier1981}
{Roddier}, F., ``{The effects of atmospheric turbulence in optical astronomy},'' {\em Progess in Optics}~{\bf 19},  281--376 (Jan. 1981).

\bibitem{Tokovinin2002}
Tokovinin, A., ``Measurement of seeing and the atmospheric time constant by differential scintillations,'' {\em Appl. Opt.}~{\bf 41},  957--964 (Feb 2002).

\bibitem{Kornilov2011}
{Kornilov}, V.~G., ``{The statistics of the photometric accuracy based on MASS data and the evaluation of high-altitude wind},'' {\em Astronomy Letters}~{\bf 37},  40--48 (Jan. 2011).

\bibitem{Kornilov2012a}
Kornilov, M.~V., ``{Estimation of vertical profiles of wind from MASS measurements},'' in [{\em Adaptive Optics Systems III}{\nolinebreak\hspace{0.1em}]},  Ellerbroek, B.~L., Marchetti, E., and V{\'e}ran, J.-P., eds.,  {\bf 8447},  84471B, International Society for Optics and Photonics, SPIE (2012).

\end{thebibliography}
\bibliographystyle{spiebib} 


{\bf Victor Kornilov} was the Head of the Laboratory of New Photometric Methods at Sternberg Astronomical Institute, Lomonosov Moscow State University, an Associate Professor at the Faculty of Physics of Lomonosov Moscow State University. He received his PhD in astrophysics in 1979 from SAI MSU. His primary research interests included atmospheric turbulence, precision photometry, and automation of astronomical equipment.

{\bf Matwey Kornilov} is a researcher at Sternberg Astronomical Institute, Lomonosov Moscow State University, and an Associate Professor at the Higher School of Economics (HSE University), where he specializes in data analysis education. He received his PhD in astrophysics in 2016 from SAI MSU. His primary research interests include atmospheric turbulence and machine learning applications in astronomy.

{\bf Boris Safonov} is a researcher at Sternberg Astronomical Institute Lomonosov Moscow State University. He received his PhD in astrophysics and stellar astronomy in SAI MSU in 2012. His primary research interests are atmospheric turbulence, methods of high angular resolution in astronomy, polarimetry, and characterization of circumstellar envelopes.

{\bf Anton Mironov} is a graduate of the Specialist Program of the Faculty
of Space Research Lomonosov Moscow State University. The final
qualifying paper is dedicated to the topic "Measuring dome turbulence
using scintillation pattern on the 2.5-m telescope at the Caucasian
mountain observatory of Sternberg Astronomical Institute (Lomonosov
Moscow State University)".

{\bf Dmitry Cheryasov} is an engineer at the Sternberg Astronomical Institute, Lomonosov Moscow State University. He received a Specialist in Astronomy degree from SAI MSU in 2014. His work focuses on the mechanical and electronic design, as well as the manufacturing, of astronomical instrumentation.

Biographies of the other authors are not available.

\end{document}